\documentclass[a4paper,fleqn,usenatbib,onecolumn]{mnras}
\usepackage[T1]{fontenc}
\usepackage{ae,aecompl}
\usepackage{graphicx}	
\usepackage{amsmath}	
\usepackage{amssymb}	

\newcommand{\be}{\begin{equation}}
\newcommand{\ee}{\end{equation}}
\newcommand{\bq}{\begin{eqnarray}}
\newcommand{\eq}{\end{eqnarray}}

\title[Forecasts of redshift drift constraints]{Forecasts of redshift drift constraints on cosmological parameters}

\author[C. S. Alves et al.]{
C. S. Alves,$^{1,2}$\thanks{E-mail: catarina.alves.18@ucl.ac.uk (CSA)}
A. C. O. Leite,$^{1,3,4}$\thanks{E-mail: Ana.Leite@astro.up.pt (ACOL)}
C. J. A. P. Martins,$^{1,3}$\thanks{E-mail: Carlos.Martins@astro.up.pt (CJAPM)}
J. G. B. Matos$^{1,3,4}$\thanks{E-mail: up201506054@fc.up.pt (JGBM)}
\newauthor and T. A. Silva$^{1,4}$\thanks{E-mail: up201405824@fc.up.pt (TAS)}
\\
$^{1}$Centro de Astrof\'{\i}sica da Universidade do Porto, Rua das Estrelas, 4150-762 Porto, Portugal\\
$^{2}$Department of Physics and Astronomy, University College London, Gower Street, London WC1E 6BT, United Kingdom\\
$^{3}$Instituto de Astrof\'{\i}sica e Ci\^encias do Espa\c co, CAUP, Rua das Estrelas, 4150-762 Porto, Portugal\\
$^{4}$Faculdade de Ci\^encias, Universidade do Porto, Rua do Campo Alegre, 4150-007 Porto, Portugal
}

\date{Accepted XXX. Received YYY; in original form ZZZ}

\pubyear{2019}

\begin{document}
\label{firstpage}
\pagerange{\pageref{firstpage}--\pageref{lastpage}}
\maketitle

\begin{abstract}
Cosmological observations usually map our present-day past light cone. However, it is also possible to compare different past light cones. This is the concept behind the redshift drift, a model-independent probe of fundamental cosmology. In simple physical terms, this effectively allows us to watch the Universe expand in real time. While current facilities only allow sensitivities several orders of magnitude worse than the expected signal, it should be possible to detect it with forthcoming ones. Here we discuss the potential impact of measurements by three such facilities: the Extremely Large Telescope (the subject of most existing redshift drift forecasts), but also the Square Kilometre Array and intensity mapping experiments. For each of these we assume the measurement sensitivities estimated respectively in Liske {\it et al.} (2008), Klockner  {\it et al.} (2015) and Yu {\it et al.} (2014). We focus on the role of these measurements in constraining dark energy scenarios, highlighting the fact that although on their own they yield comparatively weak constraints, they do probe regions of parameter space that are typically different from those probed by other experiments, as well as being redshift-dependent. Specifically, we quantify how combinations of several redshift drift measurements at different redshifts, or combinations of redshift drift measurements with those from other canonical cosmological probes, can constrain some representative dark energy models. Our conclusion is that a model-independent mapping of the expansion of the universe from redshift $z=0$ to $z=4$---a challenging but feasible goal for the next generation of astrophysical facilities---can have a significant impact on fundamental cosmology.
\end{abstract}



\section{Introduction}

The idea that the redshift of objects following the cosmological expansion changes with time---known as the redshift drift--- is many decades old, and was first coherently formulated by \citet{Sandage,Mcvittie}. The idea was revived about 20 years ago by \citet{Loeb}, who also provided a first discussion of possible astrophysical systems in which to carry out the measurements. Indeed, one often refers to redshift drift measurements using the Lyman-alpha forest as the Sandage-Loeb test, though in this article we will use the general term since we will be concerned with its measurement by several astrophysical facilities and relying on correspondingly different techniques.

Conceptually, a measurement of the redshift drift is of fundamental importance. In our usual astrophysical observations at cosmological distances done so far, we are effectively mapping our present-day past light cone. An alternative---which is possible at least in principle---is to compare different past light cones. To put it somewhat more simply, this would correspond to watching the Universe expand in real time. Other than the fact that it is, operationally, a different probe of the universe, its conceptual importance stems from the fact that it is a model-independent probe of the expansion of the universe, making no assumption on geometry, clustering or the behaviour of gravity, and therefore of crucial importance for fundamental cosmology.

The practical difficulty is simply that cosmologically relevant timescales are orders of magnitude larger than human timescales, and therefore a measurement of the redshift drift certainly requires exquisite sensitivity. Current facilities only allow sensitivities several orders of magnitude worse than the expected signal: indeed the best currently available bound, by \citet{Darling}, is about three orders of magnitude larger than the signal expected for the standard $\Lambda$CDM cosmology with reasonable choices of its model parameters, and likely to be vulnerable to systematic errors comparable to the statistical ones. Nevertheless, several previous analyses have suggested that it is possible to measure the redshift drift with forthcoming facilities, and in the case of the Extremely Large Telescope (ELT) this measurement was the key science and design driver for the development of one of its instruments, a high-resolution spectrograph currently known as ELT-HIRES \citep{HIRES}.

A detailed feasibility study of high-redshift measurements by the ELT, on a decade timescale, was done by \citet{Liske}, who also discussed possible targets. The impact of these ELT measurements for constraining cosmological models has been subsequently discussed by several authors \citep{Corasaniti,Lake,Balbi,Moraes,Geng}. The impact of the combination of ELT redshift drift measurements with Cosmic Microwave Background (CMB) data has been quantified in \cite{CMB}. An important result of these works is that although ELT redshift drift measurements, on their own, lead to cosmological parameter constraints that are not tighter than those available by more classical probes (such as supernovae or the CMB) they do probe regions of parameter space that are different from (and sometimes actually orthogonal to) those of other probes, enabling the breaking of degeneracies and therefore leading to more stringent combined constraints. In the present work we further quantify this statement. Two other recent forecasts, with more optimistic assumptions on the ELT performance, can be found in \citep{Jimenez,Lazkoz}.

More recently, it has also been pointed out by \citet{Klockner} that measurements at low redshifts can in principle be made by the Square Kilometre Array (SKA), although the full (Phase 2) SKA will be necessary to allow the measurements to be carried out in realistic amounts of time. Similarly, it has been suggested by \citet{Chime} that redshift drift measurements at intermediate redshifts can be done by intensity mapping measurements such as the Canadian Hydrogen Intensity Mapping Experiment (CHIME), and this analysis should also apply to the very similar Hydrogen Intensity and Real-time Analysis eXperiment, HIRAX \citep{HIRAX}. We note that enabling these measurements will require appropriate hardware configurations which are beyond the scope of this work; our goal here is to quantify the potential cosmological impact of realizing these measurements.

Redshift drift measurements are a key part of what is commonly called real-time cosmology---see \citet{Quercellini} for a recent review. Other possible measurement techniques have been conceptually discussed \citep{Stebbins,Kim}. Interestingly, one could even measure the redshift drift using the CMB, although the required timescale would be one century \citep{Lange}. (Note that when ordinarily looking at the CMB on sufficiently large scales one is comparing different past light cones at the same time.) All of these pertain to the first derivatives of the redshift; a generic analysis (i.e., not specific to any particular experiment) of its combination with low redshift data has been done by \citet{Neben}. The possible effects of cosmological perturbations and inhomogeneities have also been discussed in \citet{Mellier,Koksbang}, while tests of the Copernican Principle were addressed by \citet{Uzan}. More recently the role of second derivatives of the redshift, which should also be within the reach of the full SKA, has been studied by \citet{Second}.

Here we use Fisher Matrix techniques \citep{FMA1,FMA2} for a comparative study discussing the cosmological impact of redshift drift measurements by the ELT, the SKA and CHIME, on their own and in combination, and also using additional priors representative of CMB and other measurements. Together, these facilities span the redshift range from $z=0$ to beyond $z=4$, enabling a study of the dynamics of the universe from the deep matter era, through the onset of the acceleration phase and until the present epoch. Our goal is to use a common analysis methodology for all three facilities, and in particular providing forecasts for SKA and CHIME that can be compared to (and combined with) those for the ELT. For concreteness we will use three fiducial models: canonical $\Lambda$CDM, a constant dark energy equation of state ($w_0$CDM) and the well known Chevallier-Polarski-Linder (CPL) parametrization \citep{CPL1,CPL2}. Unless otherwise is stated, flat models are assumed.

\section{Theoretical formalism and statistical tools}

We start with a brief pedagogical overview, introducing the concept of redshift drift, illustrating its dependence on model parameters for our choices of fiducial models, and also recalling the Fisher Matrix techniques that will be used for the forecasts. While most of the methodology in this section is standard, we discuss it here as a simple worked example, with the goal of providing a self-contained treatment and also because it can be used to obtain some simple analytic results which are helpful to interpret the numerical results to be discussed in the latter part of the work.

\subsection{Redshift drift phenomenology}

The redshift drift of an astrophysical object following the cosmological expansion, for an observer looking at it over a time span $\Delta t$, can be shown to be given by \citep{Sandage,Liske,Second}
\be
\frac{\Delta z}{\Delta t}=H_0 \left[1+z-E(z)\right]\,,
\ee
although the actual astrophysical observable is usually a spectroscopically measured velocity
\be\label{specvel}
\Delta v=\frac{c\Delta z}{1+z}=(cH_0\Delta t)\left[1-\frac{E(z)}{1+z}\right]\,.
\ee
Note that for future convenience we have defined the rescaled Hubble parameter
\be
E(z)=\frac{H(z)}{H_0}\,,
\ee
with $H_0$ denoting the present-day value of the Hubble parameter (in other words, the Hubble constant).

The dependence on the Hubble parameter $H(z)$ naturally leads to a redshift dependence of the drift which will be model-dependent. Broadly speaking, in a universe that is currently accelerating but was deccelerating in the past the drift will be positive at low redshifts and negative for higher redshifts, while in a universe that always deccelerates the redshift drift would always be negative. It is therefore instructive to consider two specific redshifts at which the drift behaviour changes (even if in practice these redshifts can't be directly measured, but would have to be inferred from measurements at other redshifts). Apart from trivially vanishing at $z=0$, the signal will also vanish at
\be
\Delta z=0 \Longleftrightarrow \Delta v=0 \Longleftrightarrow E^2(z)=(1+z)^2\,;
\ee
naturally this zero-signal redshift will be the same for the spectroscopic velocity and the drift itself. On the other hand the redshift of maximal (positive) signal will be different for the drift and the velocity, due to the difference of the $(1+z)$ factor. Specifically, for the former we have
\be
(\Delta z)'=0 \Longleftrightarrow \frac{dE(z)}{dz}=1\,,
\ee
while for the latter
\be
(\Delta v)'=0 \Longleftrightarrow \frac{d}{dz} \left(\frac{E(z)}{1+z}\right)=0\,.
\ee

As mentioned in the introduction we will use the CPL parametrization as our most general fiducial model. In this case we can write the Friedmann equation as
\be\label{cpl}
E^2(z)=\Omega_k(1+z)^2+\Omega_m(1+z)^3+\Omega_\phi(1+z)^{3(1+w_0+w_a)}\exp{\left[-\frac{3w_az}{1+z}\right]}\,.
\ee
It follows that the redshift of zero drift is given by the solution of
\be
\Omega_m(1+z)+\Omega_\phi(1+z)^{1+3(w_0+w_a)}\exp{\left[-\frac{3w_az}{1+z}\right]}=1-\Omega_k\,,
\ee
while the redshift of maximum positive drift will be a solution of
\be
2\Omega_k(1+z)+3\Omega_m(1+z)^2+3\Omega_\phi\left[1+w_0+\frac{w_az}{1+z}\right] (1+z)^{2+3(w_0+w_a)} \exp{\left[-\frac{3w_az}{1+z}\right]} =2E(z)\,,
\ee
and the redshift of the maximum spectroscopic velocity is the solution of
\be
\Omega_m+\Omega_\phi\left[1+3w_0+\frac{3w_az}{1+z}\right] (1+z)^{3(w_0+w_a)} \exp{\left[-\frac{3w_az}{1+z}\right]} =0\,.
\ee
It is worthy of note that the first two depend on $\Omega_k$, while the last one is independent of it. For a discussion of how redshift drift measurements can help constrain curvature see \citet{Jimenez}. It's also important to bear in mind that none of these depend on the value of the Hubble constant, which simply provides an overall normalization (that is, multiplicative) factor.

As a simple illustration, for flat $\Lambda$CDM the redshift of zero drift is given by
\be
z_{zero}=\frac{1-3\Omega_m+\sqrt{1+2\Omega_m-3\Omega^2_m}}{2\Omega_m}
\ee
(where we have neglected the unphysical negative solution) while the redshift of maximum spectroscopic velocity is
\be
z_{v,max}=\left[\frac{2(1-\Omega_m)}{\Omega_m}\right]^{1/3}-1\,.
\ee
On the other hand the maximum of the drift is obtained by solving the quartic equation
\be
9\Omega_m^2(1+z)^4=4\Omega_m(1+z)^3+4(1-\Omega_m)\,.
\ee
To give a concrete example, if we choose $\Omega_m=0.3$ (in agreement with modern cosmological data) we obtain
\be
z_{zero}=\sim 2.09\,,
\ee
\be
z_{v,max}\sim 0.67\,,
\ee
while the maximum of the drift occurs at
\be
z_{z,max}\sim 0.95\,.
\ee
Note that for flat $\Lambda$CDM with $\Omega_m=0.3$, the acceleration phase does start at $z_{v,max}$, which the cosmological constant only starts dominating the Friedmann equation at
\be
z_{\Lambda}=\left[\frac{(1-\Omega_m)}{\Omega_m}\right]^{1/3}-1\sim0.33\,.
\ee

\begin{figure}
\begin{center}
\includegraphics[width=3.2in,keepaspectratio]{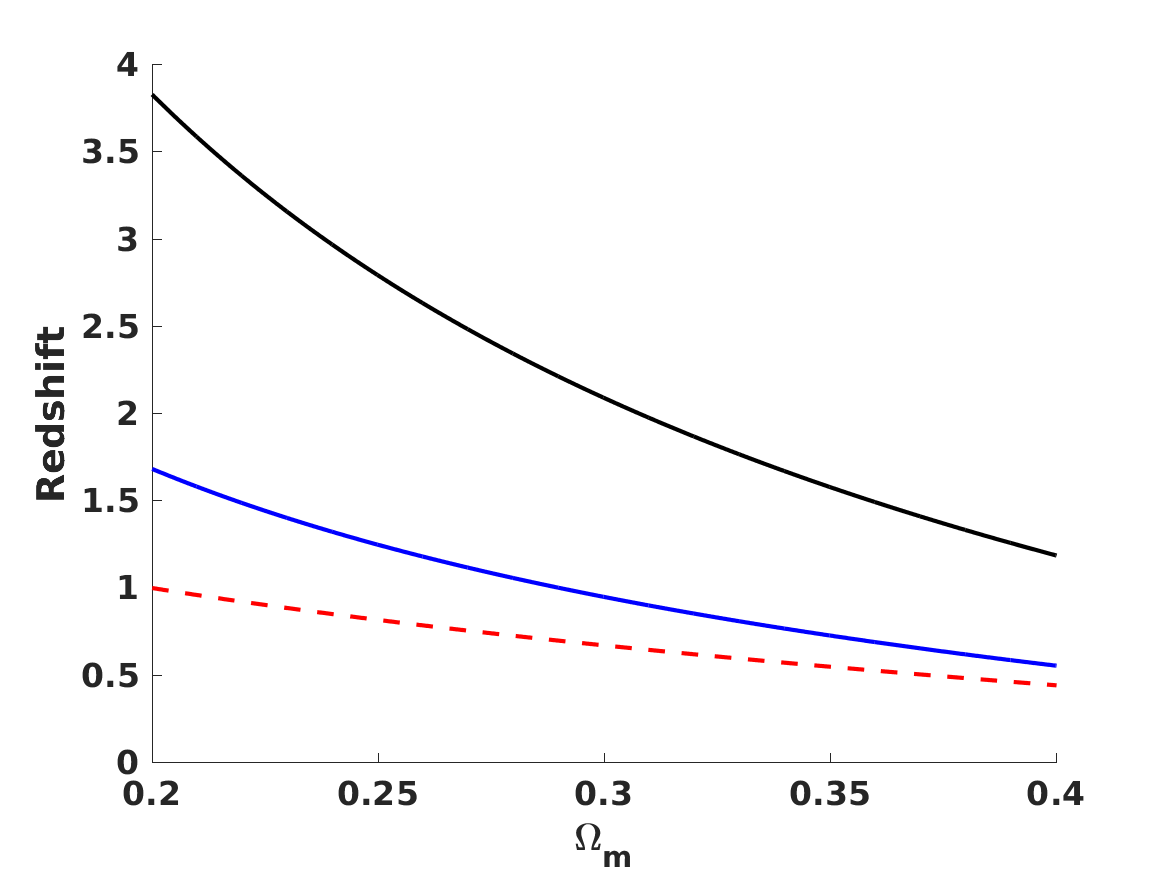}
\end{center}
\caption{The behaviour of the redshift of zero drift (black, top solid curve), the redshift of maximal drift (blue, bottom solid curve) and the redshift of maximal (positive) spectroscopic velocity (red dashed curve) as a function of the matter density, for a flat $\Lambda$CDM model.}
\label{fig1}
\end{figure}
\begin{figure*}
\begin{center}
\includegraphics[width=3.2in,keepaspectratio]{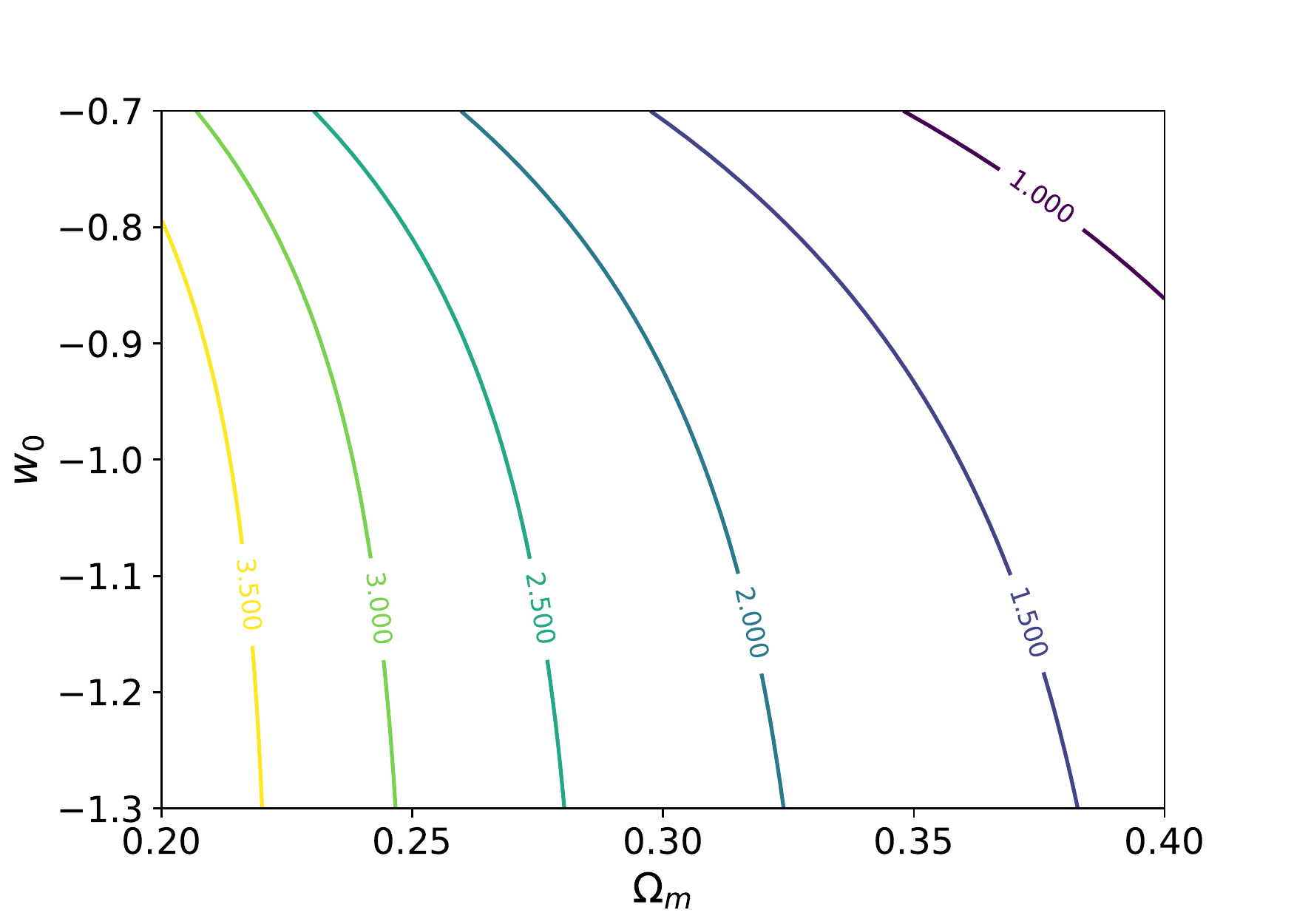}
\includegraphics[width=3.2in,keepaspectratio]{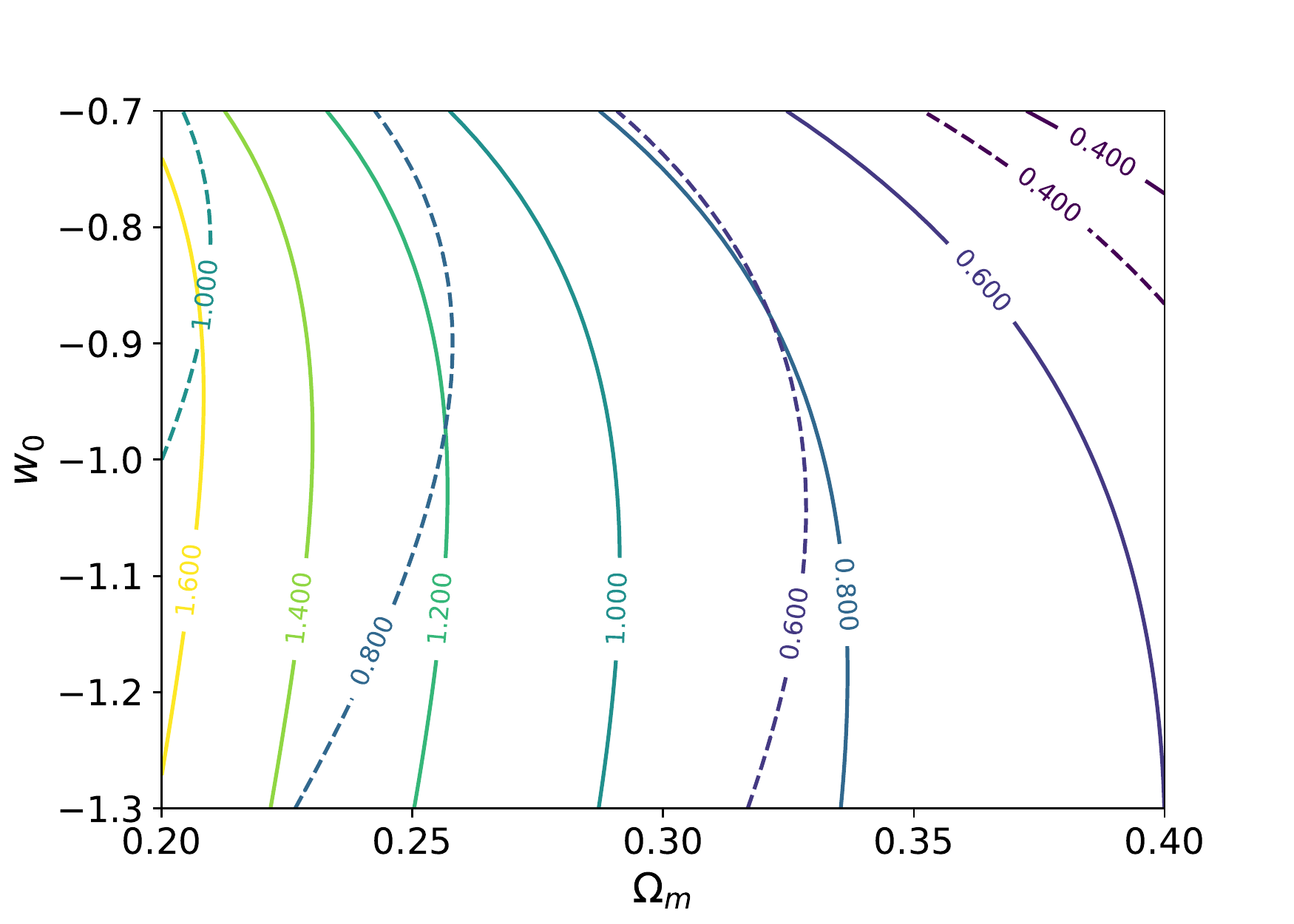}
\end{center}
\caption{The behaviour of the redshift of zero drift (left panel), the redshift of maximal drift (right panel, solid curves) and the redshift of maximal (positive) spectroscopic velocity (right panel, dashed curves) as a function of the matter density and the dark energy equation of state parameter, for a flat $w_0$CDM model.}
\label{fig2}
\end{figure*}
\begin{figure*}
\begin{center}
\includegraphics[width=3.2in,keepaspectratio]{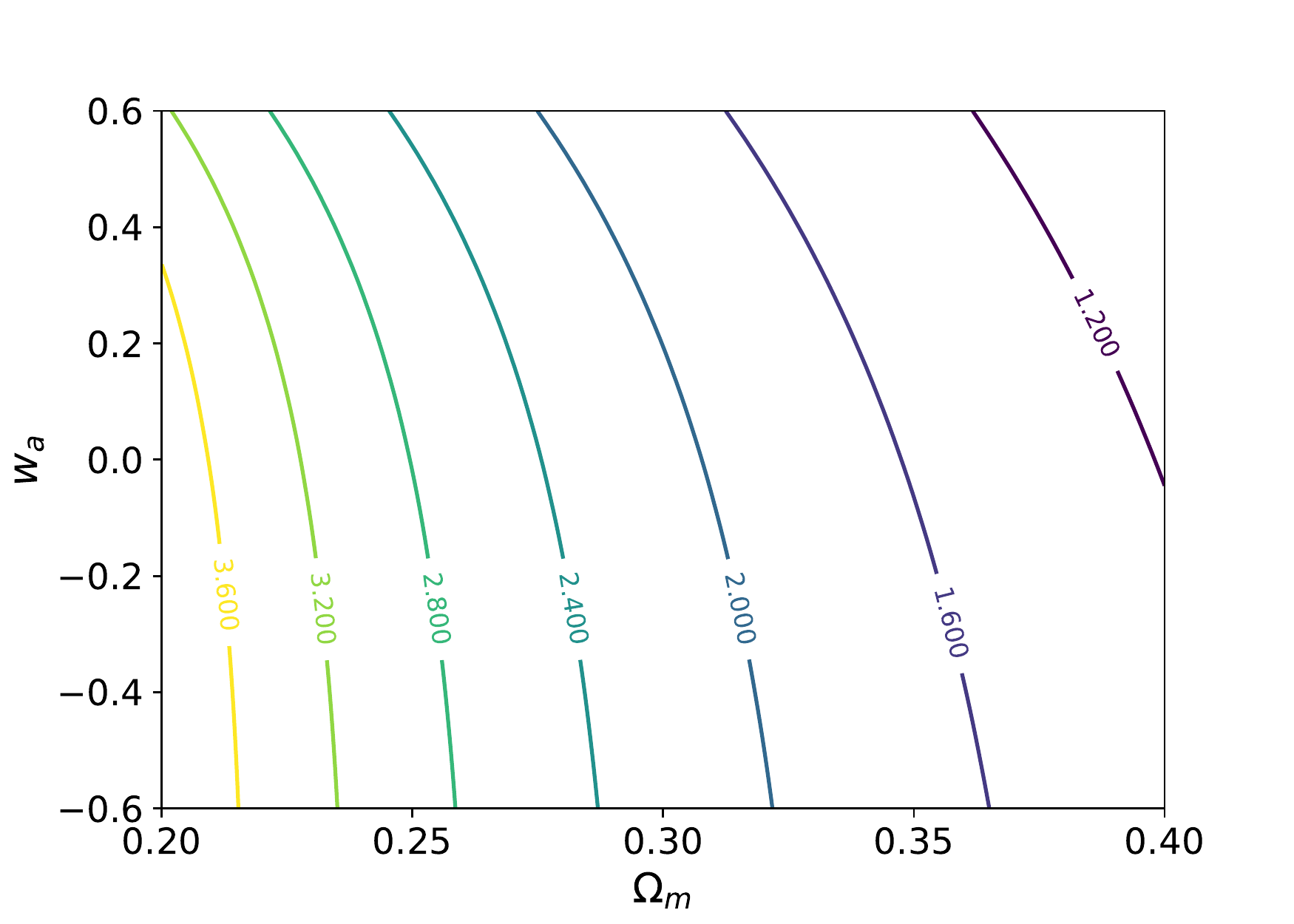}
\includegraphics[width=3.2in,keepaspectratio]{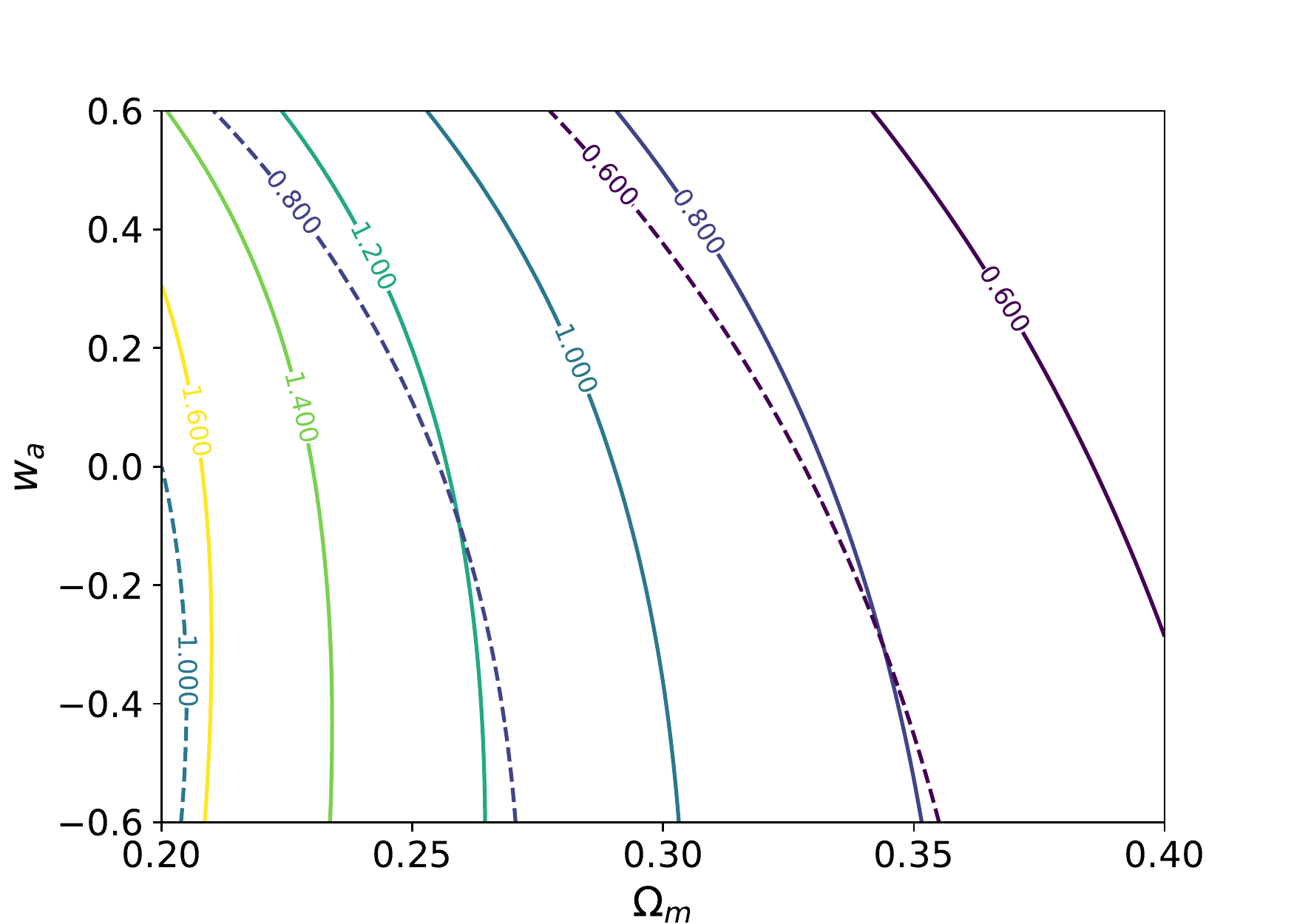}
\includegraphics[width=3.2in,keepaspectratio]{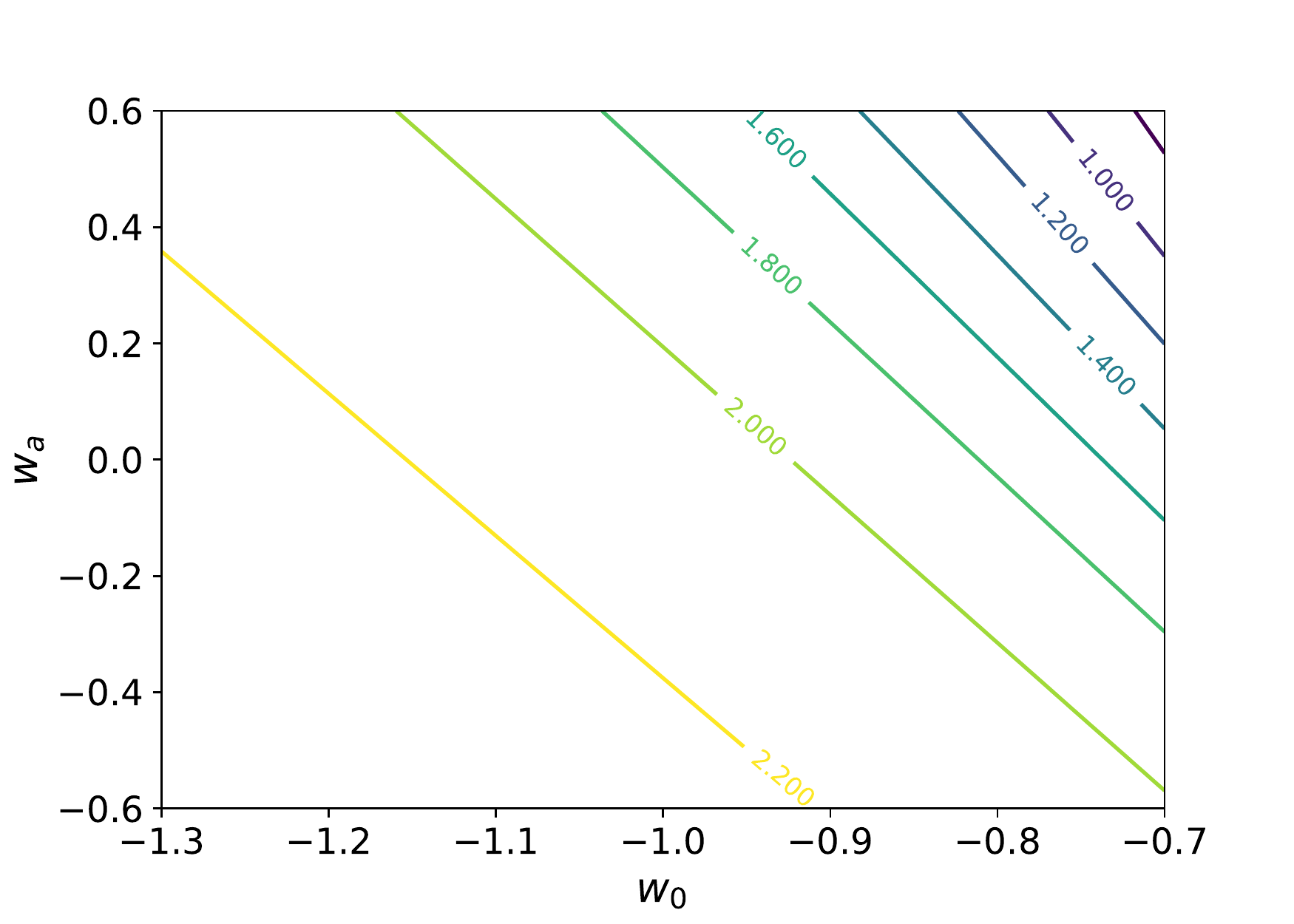}
\includegraphics[width=3.2in,keepaspectratio]{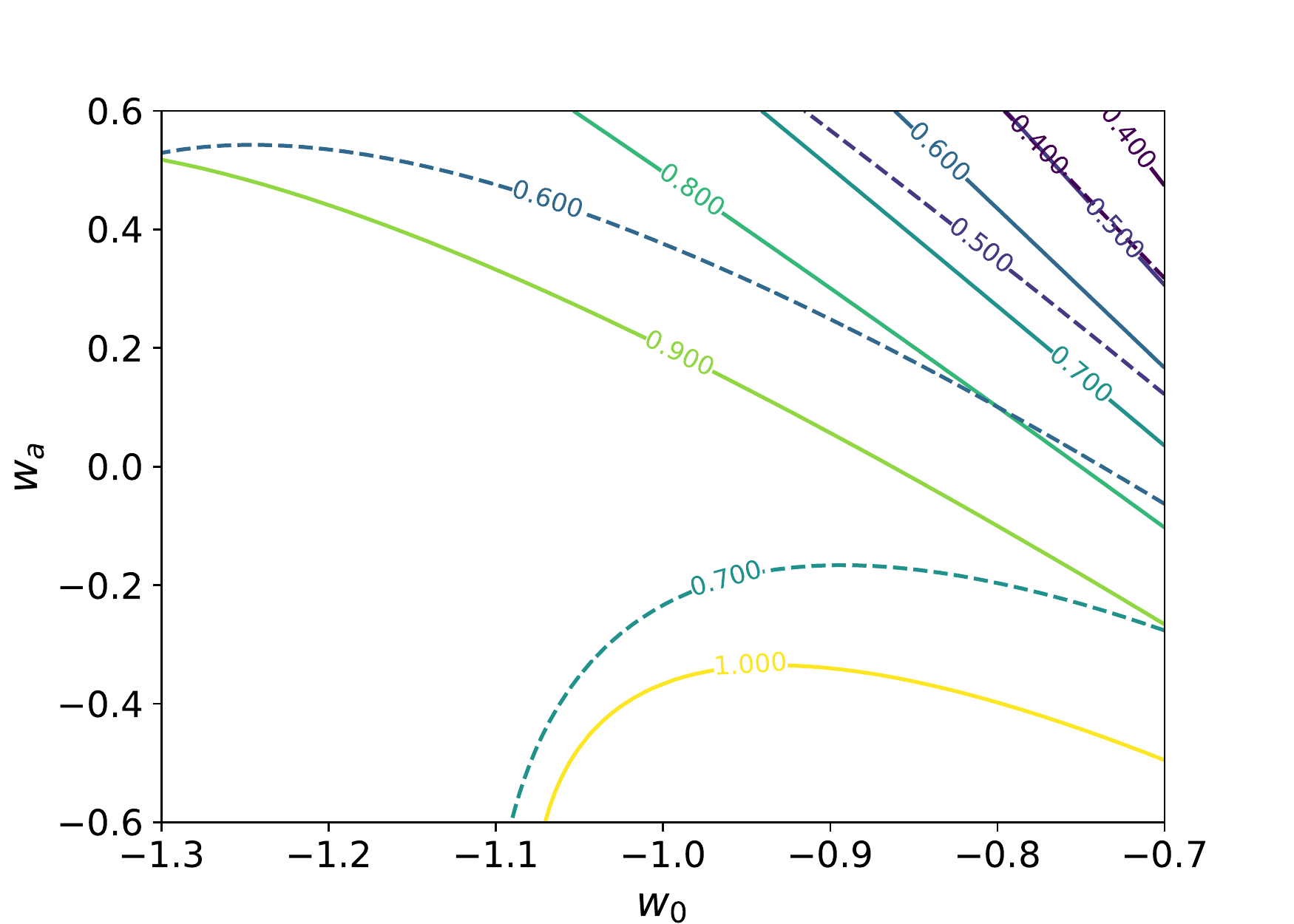}
\end{center}
\caption{The behaviour of the redshift of zero drift (left panels), the redshift of maximal drift (right panels, solid curves) and the redshift of maximal (positive) spectroscopic velocity (right panels, dashed curves) as a function of the matter density and the dark energy equation of state parameters, for a flat CPL model. In the top panels $w_0=-1$, while in the bottom panels $\Omega_m=0.3$.}
\label{fig3}
\end{figure*}

Figure \ref{fig1} illustrates the dependence of the redshifts of zero and maximal drift and velocity as a function of the matter density $\Omega_m$, for a flat $\Lambda$CDM model. Figures \ref{fig2} and \ref{fig3} show analogous contour plots for the constant equation of state model and the CPL model. As is to be expected, a smaller matter density (corresponding to a larger dark energy density, under the flatness assumption) leads to increased values for the redshifts of maximal and zero drift. On the other hand, the effect of the dark energy equation of state is less obvious. This is well illustrated in the CPL case, and can be seen in the bottom panels of Fig. \ref{fig3}: to a good approximation the redshift of zero drift depends on the sum $(w_0+w_a/3)$, but the dependence is not as simple for the maximal drift or spectroscopic velocity. These differences will be important when we discuss forecasts.

\subsection{Fisher Matrix analysis}

Our primary goal is to forecast the uncertainties with which one will be able to constrain cosmological parameters with future redshift drift measurements, from the ELT, SKA, and CHIME. Our forecasts will be done using standard Fisher Matrix analysis techniques \citep{FMA1,FMA2}, which we start by briefly summarizing for the sake of completeness. If we have a set of M model parameters $(p_1, p_2, ..., p_M)$ and the model's predictions for N observables---that is, measured quantities---$(f_1, f_2, ..., f_N)$, then the Fisher matrix is
\be
F_{ij}=\sum_{a=1}^N\frac{\partial f_a}{\partial p_i}\frac{1}{\sigma^2_a}\frac{\partial f_a}{\partial p_j}\,.
\ee
Previously known uncertainties on all (or some of) the parameters, known as priors, can be trivially added to the calculated Fisher matrix. In our case the observables are the various measurements of the spectroscopic velocity.

To be specific, in what follows let us take the case of a Fisher matrix for two model parameters, x and y. The inverse of the Fisher matrix is the covariance matrix, 
\be
[F]^{-1}\equiv [C]=\begin{bmatrix}
\sigma^2_{x} \hfill & \sigma^2_{xy} \hfill \\
\sigma^2_{xy} \hfill & \sigma^2_{y}  \hfill
\end{bmatrix}\,,
\ee
where $\sigma^2_{x}$ and $\sigma^2_{y}$ are the uncertainties in the x and y parameters marginalizing over the other parameters (that is, taking into account the uncertainties in these other parameters), while $\sigma^2_{xy}=\rho \sigma_{x}\sigma_{y}$ with $\rho$ being the correlation coefficient, ranging from $\rho=0$ for independent parameters to $\rho=\pm1$ for fully correlated and fully anticorrelated parameters. It's also useful to define a Figure of Merit
\be
FoM=\frac{1}{\sigma_x\sigma_y\sqrt{1-\rho^2}(\Delta\chi^2)}=\frac{1}{\left(\sigma^2_{x} \sigma^2_{y} - \sigma^4_{xy}  \right)^{1/2}(\Delta\chi^2)}\,,\label{FoMs}
\ee
which is proportional to the inverse of the area of the confidence ellipse of the two parameters: a small area (meaning small uncertainties in the parameters) corresponds to a large figure of merit. In the above, $\Delta\chi^2$ identifies the confidence interval of interest; for example, $\Delta\chi^2=2.3$ corresponds to the $68.3\%$ confidence level, which is the choice we used in what follows.

The above can straightforwardly be generalized to more than two parameters. If we start with a larger parameter space and later want to reduce it and calculate a new Fisher matrix marginalized over any variable we simply remove that variable's row and column from the covariance matrix and then calculate its inverse to find the new Fisher matrix. Conversely, if we want to assume perfect knowledge of a parameter, we remove that parameter's row and column from the Fisher matrix and invert it to get the new covariance matrix and parameter uncertainties.

As an illustration we will start by calculating the Fisher matrix analytically for a few simple examples in the next section. These will also serve as validating test of our generic pipeline, which has been implemented numerically. For the moment we will also analytically explore the sensitivity of the redshift drift to the cosmological parameters, for a fiducial flat CPL model---which will prove useful in the interpretation of our subsequent results. In this case, and restricting ourselves to flat models ($\Omega_k=0$) we have
\be
E^2(z)=\frac{H^2(z)}{H_0^2}=\Omega_m(1+z)^3+(1-\Omega_m)(1+z)^{3(1+w_0+w_a)}\exp{\left[\frac{-3w_az}{1+z}\right]}\,.
\ee
It is also convenient to define a dimensionless redshift drift
\be
S_z=\frac{1}{H_{100}}\frac{\Delta z}{\Delta t}=h\left[1+z-E(z)\right]\,,
\ee
where we have defined $H_0=hH_{100}$ and $H_{100}=100$ km/s/Mpc; the corresponding observable spectroscopic velocity will be denoted
\be
S_v={\Delta v}=kh\left[1-\frac{E(z)}{1+z}\right]\,.
\ee
In these we have also introduced $k=cH_{100}\delta t$, which is a constant parameter, for a given observation time, with units of cm/s. (Do not confuse this $k$ with the curvature parameter, which is set to zero in everything that follows.) Specifically for $\delta t=1$ year we have $k=3.064$ cm/s. This provides a rough estimate of the magnitude of the redshift drift in the relevant time span, and therefore also an estimate of the required sensitivity of the spectroscopic measurements. Then for each cosmological parameter $p_i$
\be
\frac{{\partial S_v}/{\partial p_i}}{{\partial S_z}/{\partial p_i}}=\frac{k}{1+z}
\ee
For our fiducial flat $\Lambda$CDM model we have
\be
\frac{\partial S_z}{\partial h}=1+z-E(z)
\ee
\be
\frac{\partial S_z}{\partial\Omega_m}=-\frac{h(1+z)^3}{2E(z)}\left[1-(1+z)^{3(w_0+w_a)}\exp{\left[\frac{-3w_az}{1+z}\right]}\right]
\ee
\be
\frac{\partial S_z}{\partial w_0}=-\frac{3h(1-\Omega_m)}{2E(z)}(1+z)^{3(1+w_0+w_a)}\ln{(1+z)}\exp{\left[\frac{-3w_az}{1+z}\right]}
\ee
\be
\frac{\partial S_z}{\partial w_a}=-\frac{3h(1-\Omega_m)}{2E(z)}(1+z)^{3(1+w_0+w_a)}\left[\ln{(1+z)}-\frac{z}{1+z}\right]\exp{\left[\frac{-3w_az}{1+z}\right]}
\ee
Note that the sign of the ${\partial S_z}/{\partial h}$ term will depend on redshift, while those of the other derivatives are always negative for observationally reasonable values of the model parameters.

Some sensitivity ratios are illuminating. Starting with the two dark energy equation of state parameters
\be
\frac{{\partial S_z}/{\partial w_a}}{{\partial S_z}/{\partial w_0}}=1-\frac{z}{(1+z)\ln{(1+z)}}\,,
\ee
and as one would expect this tends to zero as $z\longrightarrow 0$ and to unity as $z\longrightarrow \infty$. On the other hand, comparing the sensitivities to the matter density and the present-day dark energy equation of state one finds
\be
\frac{{\partial S_z}/{\partial \Omega_m}}{{\partial S_z}/{\partial w_0}}=\frac{1-(1+z)^{3(w_0+w_a)}\exp{\left[\frac{-3w_az}{1+z}\right]}}{3(1-\Omega_m)\ln{(1+z)}(1+z)^{3(w_0+w_a)}\exp{\left[\frac{-3w_az}{1+z}\right]}}
\ee
which tends to
\be
\frac{-w_0}{1-\Omega_m}\,,\quad z\longrightarrow 0
\ee
and again to infinity as $z\longrightarrow \infty$ (though note that in this we are neglecting the radiation density). Specifically for a fiducial $\Lambda$CDM model with $\Omega_m=0.3$, at $z=1$, the ratios are respectively
\be
\frac{{\partial S_z}/{\partial w_a}}{{\partial S_z}/{\partial w_0}}=1-\frac{1}{2\ln{2}}\sim0.28
\ee
\be
\frac{{\partial S_z}/{\partial \Omega_m}}{{\partial S_z}/{\partial w_0}}\sim\frac{7}{2.1\ln{2}}\sim4.8\,.
\ee

It's also illuminating to consider the low-redshift limits of the various derivative terms. To obtain them, one uses the fact that as $z\longrightarrow 0$ we have
\be
E(z)=1+\frac{3}{2}\left[1+(1-\Omega_m)w_0\right]z+{\cal O}(z^2)\,.
\ee
For the derivatives with respect to $h$, $\Omega_m$ and $w_0$ we then find, respectively
\be
\frac{\partial S_z}{\partial h}\longrightarrow -\frac{1}{2}\left[1+3(1-\Omega_m)w_0\right]z
\ee
\be
\frac{\partial S_z}{\partial\Omega_m} \longrightarrow \frac{3}{2}hw_0 z
\ee
\be
\frac{\partial S_z}{\partial w_0} \longrightarrow -\frac{3}{2}h(1-\Omega_m)z
\ee
while that with respect to $w_a$ (whose value, interestingly, does not affect any of them) is of higher order.

\begin{figure*}
\begin{center}
\includegraphics[width=3.2in,keepaspectratio]{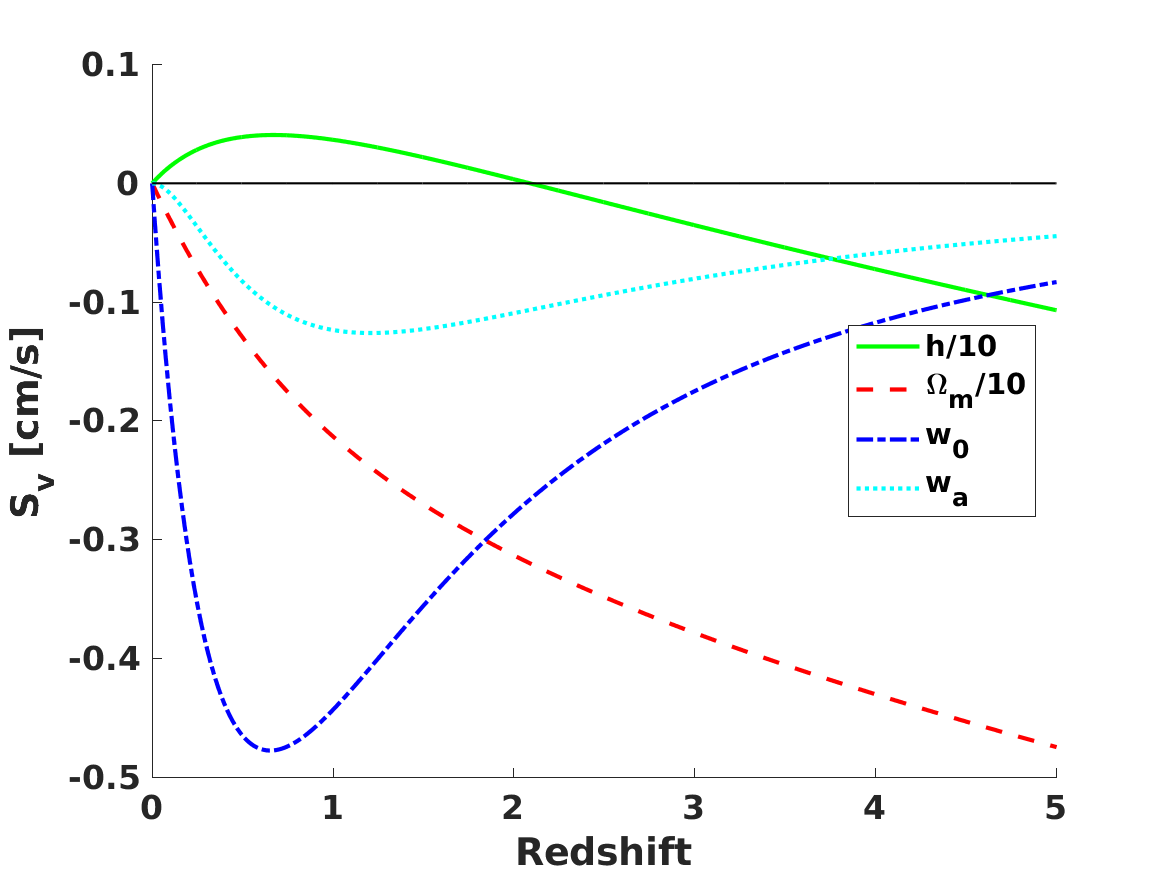}
\includegraphics[width=3.2in,keepaspectratio]{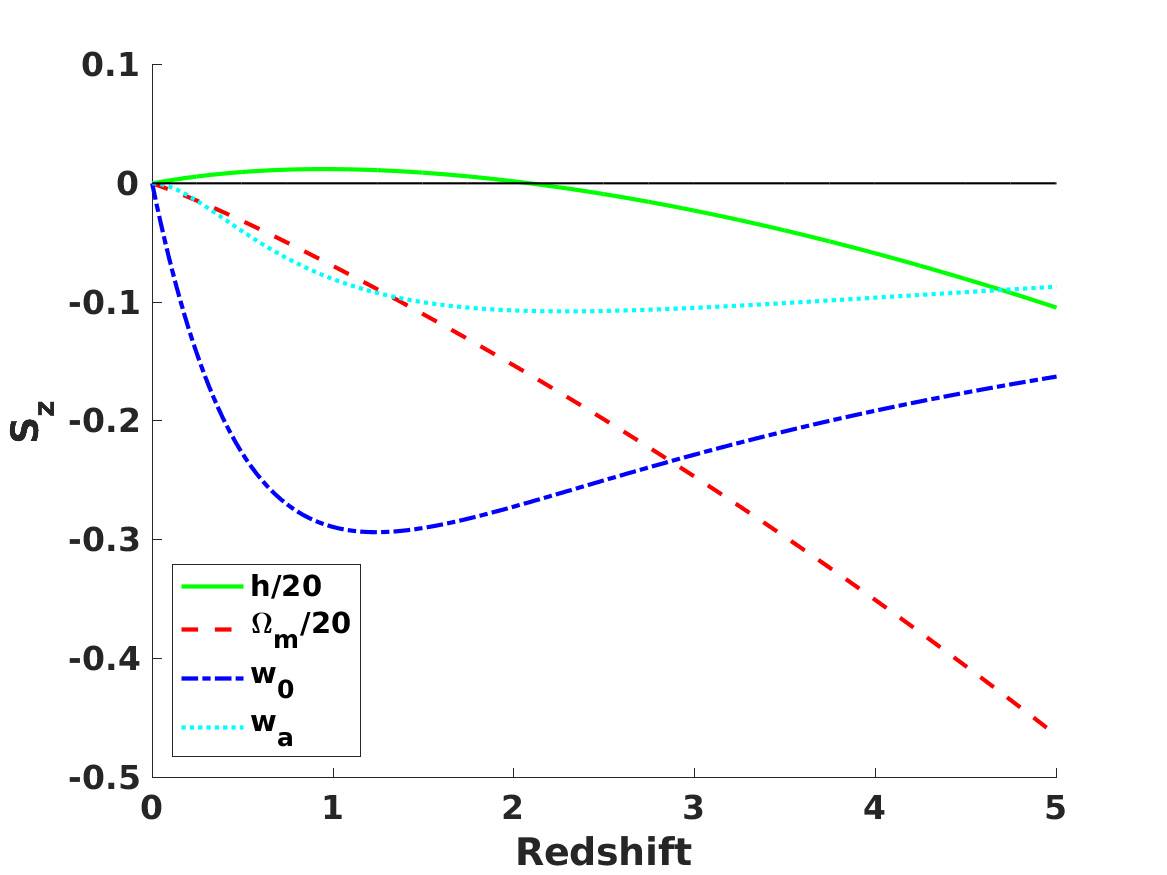}
\end{center}
\caption{Sensitivities of the spectroscopic velocity (left panel, in units of cm/s) and the redshift drift (right panel, in dimensionless units) to the cosmological parameters in the CPL parametrization, for a flat $\Lambda$CDM fiducial. Note that for plotting convenience the sensitivities to $h$ and $\Omega_m$ have been divided by a factor of 10 in the left panel and by a factor of 20 in the right panel. The zero sensitivity line is also shown.}
\label{fig4}
\end{figure*}

Figure \ref{fig4} depicts the redshift dependence of these observational sensitivities, $\partial S_v/\partial p_i$ as well as the corresponding $\partial S_z/\partial p_i$, for a flat $\Lambda$CDM fiducial model with $\Omega_m=0.3$ and $h=0.7$; the two are slightly different, but are of course related by a redshift-dependent factor. (An earlier analysis of the latter is in \citet{Kim}.) Also note the different units in the vertical axes of the two panels: $\partial S_z/\partial p_i$ is in dimensionless units while $\partial S_v/\partial p_i$ is shown in units of cm/s.

As expected the sensitivity on $h$ (which is proportional to the redshift drift itself) is the only one that changes sign; this and the sensitivity on $\Omega_m$ are also larger than those on the dark energy equation of state parameters $w_0$ and $w_a$. Nevertheless, the most noteworthy point is that the various sensitivity curves have different redshift dependencies. While the matter density sensitivity increases with redshift, those of $w_0$ and $w_a$ are maximal at around the onset of acceleration (just below $z=1$, with the value being slightly different for $S_v$ and $S_z$), and the former has a stronger redshift dependence than the latter. This is important because it implies that as long as one is able to do these measurements at sufficiently broad redshift ranges there should not be strong covariances between the parameters. We will further quantify this statement in what follows.

\section{A simple worked example}

Before proceeding with our general analysis it is illuminating to discuss the simple case of the flat $\Lambda$CDM model, in which case the only two free parameters are $h$ and $\Omega_m$, and a full analytic treatment is straightforward. Therefore in what follows we will consider some examples which will be relevant for the discussion in the subsequent sections.

\subsection{A single measurement with prior(s)}

Let us start by assuming that we have a single measurement of the redshift drift, at some generic redshift $z$ and with an uncertainty $\sigma_z$. Naturally in this case the two free parameters can only be separately constrained if we have some priors on at least one of them. Including priors on both the rescaled Hubble constant $h$ and the matter density $\Omega_m$, with the uncertainties being denoted respectively $\sigma_h$ and $\sigma_m$, the Fisher matrix is
\be
[F(h,\Omega_m)]=\begin{bmatrix}
\frac{k^2}{\sigma_z^2}L^2+\frac{1}{\sigma^2_h} \hfill & \frac{hk^2}{\sigma_z^2}LM \hfill \\
\frac{hk^2}{\sigma_z^2}LM \hfill & \frac{k^2}{\sigma_z^2}h^2M^2 +\frac{1}{\sigma^2_m} \hfill
\end{bmatrix}\,,
\ee
where for convenience we have defined the two functions
\be
L\equiv L(\Omega_m,z)=1-\frac{\sqrt{\Omega_m(1+z)^3+1-\Omega_m}}{1+z}
\ee
\be
M\equiv M(\Omega_m,z)= - \frac{(1+z)^2-(1+z)^{-1}}{2\sqrt{\Omega_m(1+z)^3+1-\Omega_m}}\,,
\ee
and also suppressed their explicit dependencies on the redshift and the matter density in the above Fisher matrix (and also in the calculations that follow). The un-marginalized uncertainties are
\be
\theta_h=\frac{\sigma_h\sigma_z}{\sqrt{\sigma^2_z+k^2L^2\sigma_h^2}}
\ee
\be
\theta_m=\frac{\sigma_m\sigma_z}{\sqrt{\sigma^2_z+k^2M^2h^2\sigma_m^2}}
\ee
while the matrix determinant is
\be
det F=\frac{k^2}{\sigma^2_z}\left[\frac{L^2}{\sigma_m^2}+\frac{h^2M^2}{\sigma_h^2} \right]+\frac{1}{\sigma^2_h\sigma^2_m}\,;
\ee
this would be zero in the absence of priors---as expected---but naturally existing cosmological data do provide us with these priors.

\begin{figure}
\begin{center}
\includegraphics[width=3.2in,keepaspectratio]{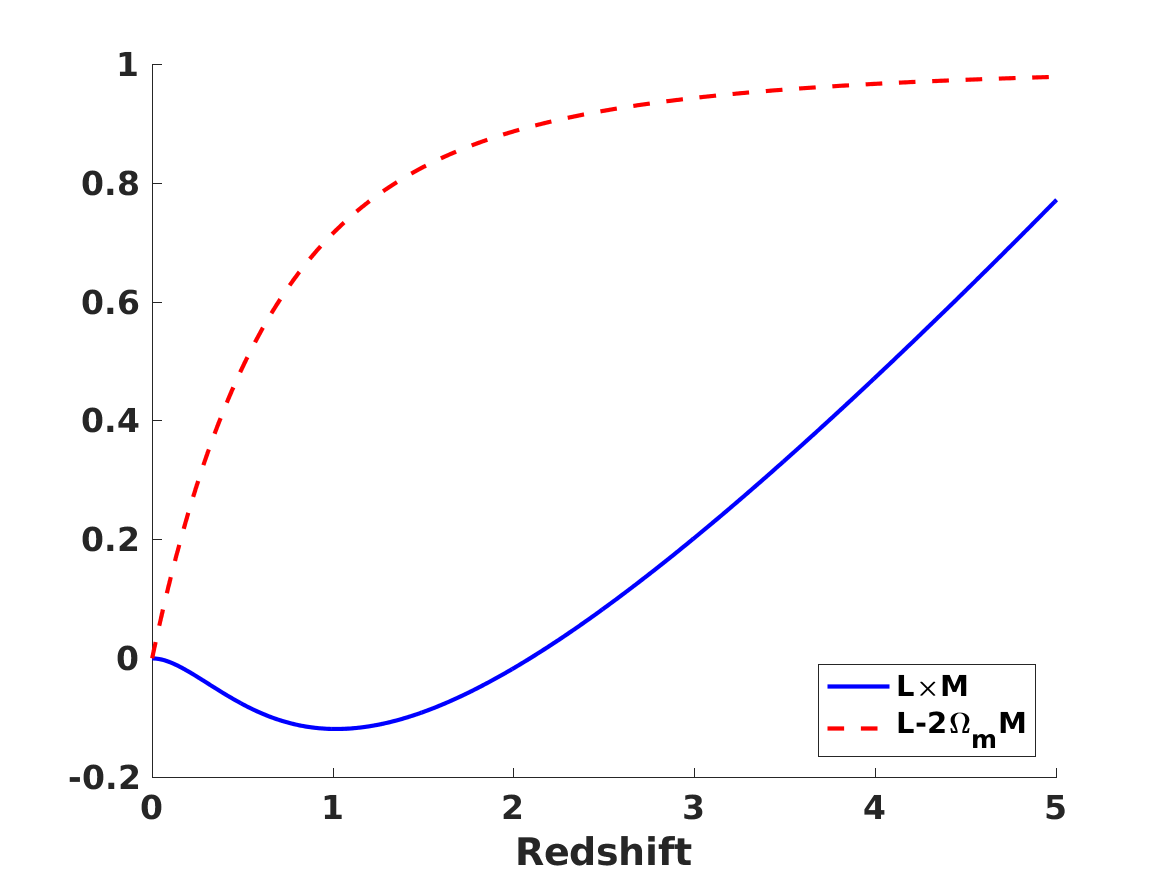}
\end{center}
\caption{Redshift dependencies of the functions $L\times M$ (blue solid line) and $L-2\Omega_mM$ (red dashed line), defined in the main text, for a matter density $\Omega_m=0.3$.}
\label{fig5}
\end{figure}

The covariance matrix is
\be
[C(h,\Omega_m)]=\frac{1}{det F}\begin{bmatrix}
\frac{k^2}{\sigma_z^2}h^2M^2 +\frac{1}{\sigma^2_m} \hfill & - \frac{hk^2}{\sigma_z^2}LM \hfill \\
- \frac{hk^2}{\sigma_z^2} LM \hfill &  \frac{k^2}{\sigma_z^2}L^2+\frac{1}{\sigma^2_h} \hfill
\end{bmatrix}\,,
\ee
and the general marginalized uncertainties are
\be
\frac{1}{\sigma^2_{h,new}}=\frac{1}{\sigma^2_h}+\frac{k^2L^2}{k^2h^2M^2\sigma^2_m+\sigma^2_z}
\ee
\be
\frac{1}{\sigma^2_{m,new}}=\frac{1}{\sigma^2_m}+\frac{k^2h^2M^2}{k^2L^2\sigma^2_h+\sigma^2_z}\,.
\ee
Finally the correlation coefficient is
\be
\rho=\, - \, \frac{LM}{ \left[L^2+\frac{\sigma^2_z}{k^2\sigma^2_h}\right]^{1/2} \left[M^2+\frac{\sigma^2_z}{k^2h^2\sigma^2_m}\right]^{1/2}} \,.
\ee
It is interesting to note that the product of the two functions
\be
LM=\frac{1}{2}\left[(1+z)^2-\frac{1}{1+z}\right]\left[\frac{1}{1+z}-\frac{1}{E(z)} \right]\,,
\ee
(in other words, the product of the first derivatives of the redshift drift with respect to each of the cosmological parameters) which determines the sign of the correlation coefficient, has a redshift-dependent sign, being negative at low redshifts and positive at high redshifts. This is depicted by the solid curve in Fig. \ref{fig5}. The redshift at which the sign changes is precisely the redshift of zero drift, discussed in the previous section. On the other hand, in the low redshift limit the product behaves as
\be
LM\sim-\frac{3}{2}\left(1-\frac{3}{2}\Omega_m\right)z^2\,,
\ee
(note that an $\Omega_m=2/3$ universe only starts accelerating today) while at high redshifts it grows linearly with redshift. Therefore in the limit where there are no priors the two parameters will be fully correlated in the case of a low-redshift measurement but fully anticorrelated in the case of a high-redshift measurement. This property is consistent with our sensitivity analysis in the precious section, and can also be seen in practice in Fig. 16 of \citet{Liske}. This will also be important for some of the subsequent discussion.

It is also interesting to contrast this with the case where one has a single prior on a combination of the two parameters, specifically on the physical matter density $\omega_m=\Omega_mh^2$. This may naturally come, for example, from CMB experiments, and we denote the corresponding uncertainly by $\sigma_\omega$. In this case the Fisher matrix has the form
\be
[F(h,\Omega_m)]=\begin{bmatrix}
\frac{k^2}{\sigma_z^2}L^2+\frac{4\Omega_m^2h^2}{\sigma^2_\omega} \hfill & \frac{hk^2}{\sigma_z^2}LM+\frac{2\Omega_mh^3}{\sigma^2_\omega} \hfill \\
\frac{hk^2}{\sigma_z^2}LM \hfill+\frac{2\Omega_mh^3}{\sigma^2_\omega} & \frac{k^2}{\sigma_z^2}h^2M^2 +\frac{h^4}{\sigma^2_\omega} \hfill
\end{bmatrix}\,,
\ee
whose determinant is
\be
det F=\frac{k^2h^4}{\sigma^2_z\sigma^2_\omega}\left(L-2\Omega_mM\right)^2\,.
\ee
It is easy to see that for observationally realistic values of the matter density the function $(L-2\Omega_mM)$ vanishes at $z=0$ but is otherwise positive; its behaviour for the case $\Omega_m=0.3$ is depicted by the dashed line in Fig. \ref{fig5}.

Finally, the non-diagonal term in the covariance matrix is
\be
\sigma_{hm}=\, -\, \frac{2\omega_m\sigma^2_z+k^2LM\sigma^2_\omega}{kh^3(L-2\Omega_mM)^2}\,.
\ee
Therefore in this case the correlation coefficient will always be positive at high redshifts (as before), but for low-redshifts the behaviour will depend on the relative values of the prior and the uncertainty of the redshift drift measurement. Specifically, a sufficiently precise measurement (that is, a sufficiently small $\sigma_z$) will be necessary to ensure a positive correlation at low redshift. The redshift at which the sign changes is no longer the the redshift of zero drift, but will depend on the values of $\sigma_z$ and $\sigma_\omega$.


\subsection{Several measurements without priors}

Let us now consider the case where we have two measurements of the redshift drift at different redshifts, but no external priors. We will extend the simplifying notation of the previous sub-section, with $L_i$ and $M_i$ now denoting the values of the corresponding functions at the two redshifts of the measurements, $z_i$. For simplicity we will assume that both measurements have the same uncertainty, which we therefore still denote $\sigma_z$.

\begin{figure*}
\begin{center}
\includegraphics[width=3.2in,keepaspectratio]{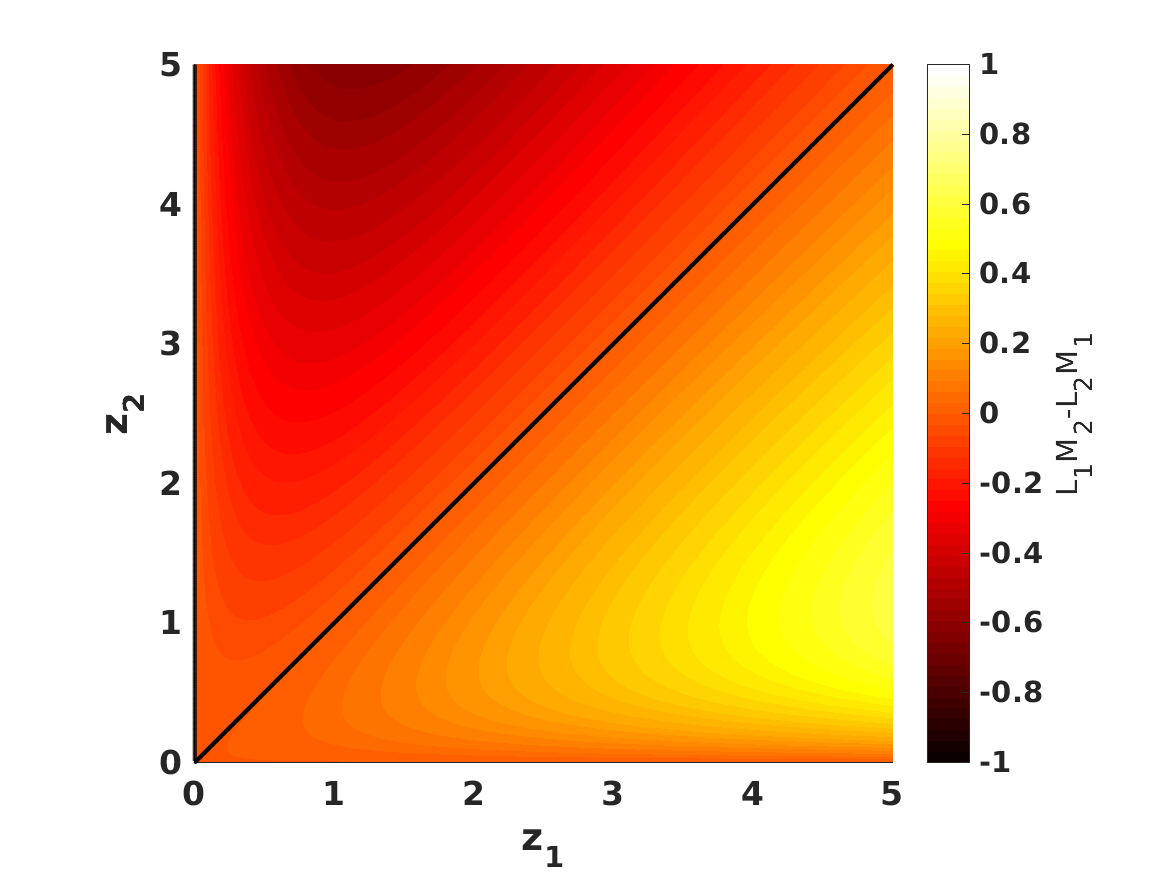}
\includegraphics[width=3.2in,keepaspectratio]{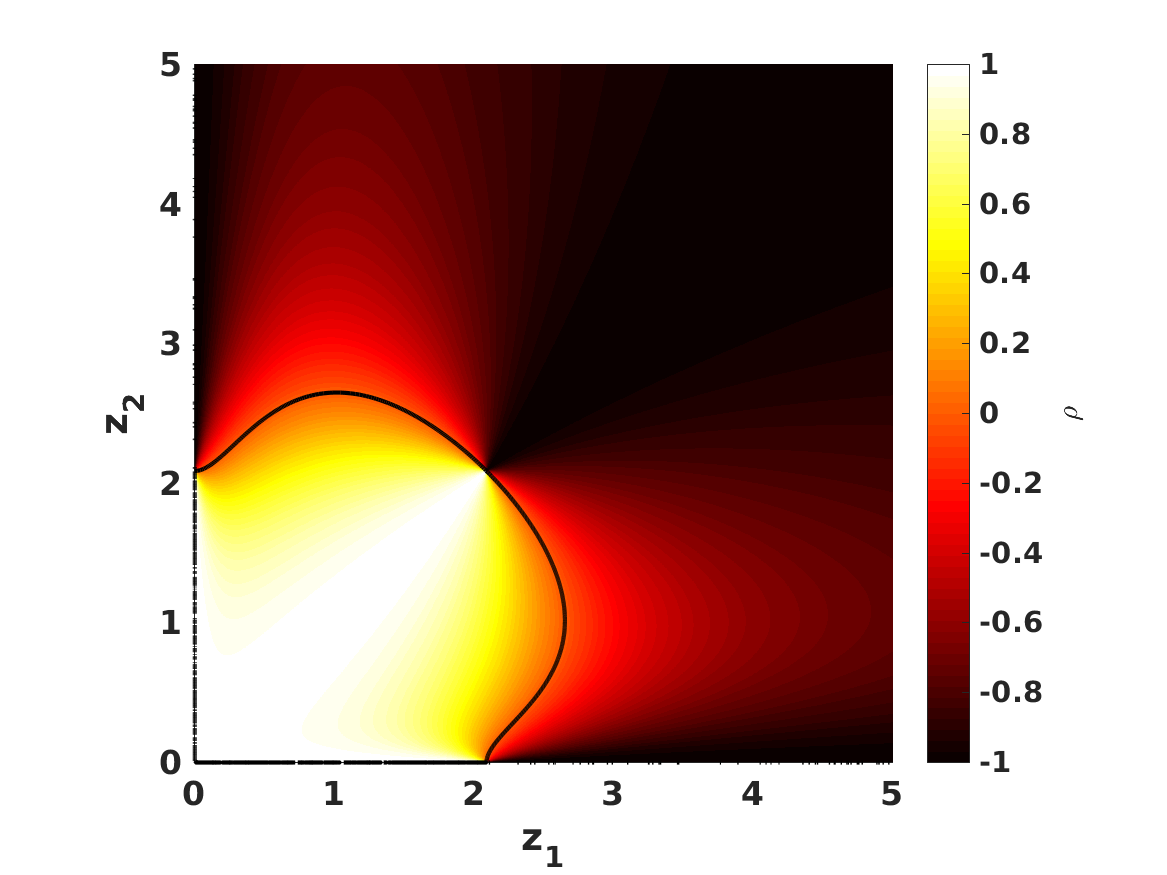}
\end{center}
\caption{Behaviour of the parameter $L_1M_2-L_2M_1$ defined in the main text (left panel) and of the correlation coefficient $\rho$ (right panel) for the case of two redshift drift measurements at redshifts $z_1$ and $z_2$, with equal uncertainties. A matter density $\Omega_m=0.3$ was also assumed. In both panels the black solid lines show where the quantity being plotted has a value of zero.}
\label{fig6}
\end{figure*}

In this case the Fisher matrix has the form
\be
[F(h,\Omega_m)]=\begin{bmatrix}
\frac{k^2}{\sigma_z^2} (L_1^2+L_2^2) \hfill & \frac{hk^2}{\sigma_z^2} (L_1M_1+L_2M_2)  \hfill \\
\frac{hk^2}{\sigma_z^2} (L_1M_1+L_2M_2)  \hfill & \frac{k^2}{\sigma_z^2}h^2(M_1^2+M_2^2) \hfill
\end{bmatrix}\,,
\ee
and the determinant is
\be
det F=\frac{k^4h^2}{\sigma^4_z}(L_1M_2-L_2M_1)^2 \,,
\ee
which only vanishes in the special cases $z_i=0$ or $z_1=z_2$---see the left panel in Fig. \ref{fig6}. Now the covariance matrix is
\be
[C(h,\Omega_m)]=\frac{\sigma^2_z}{k^2h^2\left(L_1M_2-L_2M_1\right)^2}\begin{bmatrix}
h^2(M_1^2+M_2^2) \hfill & -h(L_1M_1+L_2M_2) \hfill \\
- h(L_1M_1+L_2M_2) \hfill &  (L_1^2+L_2^2) \hfill \end{bmatrix}\,,
\ee
and the correlation coefficient is
\be
\rho=-\frac{L_1M_1+L_2M_2}{\sqrt{(L_1^2+L_2^2)(M_1^2+M_2^2)}}\,;
\ee
note that this confirms our previous results: for a single measurement without external priors the parameters are fully correlated or fully anticorrelated, depending on the redshift.

Since we have assumed that both measurements have the same uncertainty, this uncertainty cancels out from the correlation coefficient, whose generic behaviour can therefore be conveniently visualized---as is done in the right panel of Fig. \ref{fig6}. As expected a pair of low-redshift measurements will lead to correlated parameters, while a pair of high-redshift ones will yield anticorrelated parameters. Particularly stringent constraints can be obtained by combining two measurements at redshifts $z_1\sim1.0$ and $z_2=2.5$, or alternatively at $z_1\sim0.1$ and $z_2\sim2.0$. Although these numbers do depend on the relative sensitivities of the two measurements (as well as on the value of $\Omega_m$), it is interesting to note that these redshift ranges are within the observational reach, respectively of the SKA and the ELT, as we will discuss in the next section.

Finally, we also note that this example can easily be extended for the case of more than two measurements; for example for measurements at three different redshifts one finds
\be
\frac{\sigma^4_z (det F)}{k^4h^2} = (L_1M_2-L_2M_1)^2  + (L_1M_3-L_3M_1)^2 + (L_2M_3-L_3M_2)^2 \,.
\ee


\section{Forecasts for forthcoming facilities}

We are now ready to discuss forecasts for redshift drift measurements carried out by the three previously mentioned facilities: the ELT, the SKA and CHIME (or analogous intensity mapping experiments). Before doing so, we start by specifying the assumptions we make about each of them.

For the ELT, a detailed study of this science case has been done by \citet{Liske}, who found that the spectroscopic velocity uncertainty is well approximated by the following expression
\be
\sigma_v=1.35\left(\frac{S/N}{2370}\right)^{-1}\left(\frac{N_{QSO}}{30}\right)^{-1/2}\left(\frac{1+z_{QSO}}{5}\right)^{-\lambda} cm/s\,,
\ee
where the last exponent is $\lambda=1.7$ up to $z=4$ and $\lambda=0.9$ for $z>4$. Consistently with this work and other recent studies, we make the assumption of five redshift drift measurements at effective redshifts $z=2.0, 2.5, 3.0, 3.5, 4.5$ with a time span of $\Delta t=20$ years, and each with a signal to noise ratio of $S/N=3000$ and using data from $N_{QSO}=6$ quasars. These choice of time span is consistent with the latest top-level requirements for the ELT-HIRES spectrograph \citep{HIRES}, which specify a required instrument lifetime of 10 years with a goal of 20 years. The main bottleneck to these measurements (in addition to the stability of the spectrograph, which is presently understood not to be a limiting factor) is the availability of sufficiently bright quasars able to provide the required signal to noise in reasonable amounts of telescope time. Specifically, the described measurements would require a very significant amount of time with currently known quasars, but the discovery of additional bright quasars will make such an observational program more feasible, and also reduce the time span between the two epochs of observation. Such a discovery is a plausible scenario, given that the southern hemisphere of the sky is not as well explored as the northern hemisphere, and searches for additional targets are ongoing.

Admittedly, for the SKA and CHIME the feasibility of these measurements and the precision that can be achieved have been studied in much less detail. In fact, for the SKA the hardware configuration is currently not known in sufficient detail to allow a fully realistic simulation of this science case. In the case of CHIME a preliminary study has been done, though with some simplifications and without detailed simulations. Nevertheless, we will rely on currently published studies for the two facilities, while cautioning the reader that the assumptions for both of them may be somewhat optimistic (and in particular, both facilities will require suitable hardware configurations). In any case, our work provides an assessment of the impact of measurements obtained under the recently proposed scenarios detailed in the next paragraph. As previously noted, the discussion on CHIME should also be applicable to HIRAX \citep{HIRAX}.

That being said, for the SKA we will follow \citet{Klockner} in assuming five measurements at $z=0.1, 0.2, 0.3, 0.4, 0.5$ with uncertainties respectively of $2, 4, 6, 8, 10$ percent and a time span of $\Delta t=0.5$ years. Note that when referring to the SKA we always mean the full SKA, otherwise known as SKA Phase 2. It is well understood that a redshift drift measurement is unfeasible with the earlier SKA Phase 1, as it would require a time span of about 40 years. Finally, for CHIME we follow \citet{Chime} in assuming four measurements at redshifts $z=1.0, 1.4, 1.9, 2.3$, with a time span of 10 years and spectroscopic velocity uncertainties in each bin of $\sigma_v=0.8, 0.9, 1.3, 1.4$ cm/s respectively.

We will use as fiducials the three models mentioned in the introduction: $\Lambda$CDM, $w_0$CDM, and CPL, always assuming flatness. Our baseline scenario will be $\Lambda$CDM, with the relevant cosmological parameters being $\Omega_m=0.3$, $h=0.7$ and $w_0=-1$. In the case of the CPL model, in addition to the $\Lambda$CDM case ($w_0=-1, w_a=0$) we will also study two other fiducial models, ($w_0=-0.9, w_a=+0.1$) and ($w_0=-0.9, w_a=-0.1$), corresponding to freezing and thawing models in the phenomenological classification of \citet{Caldwell}.

When using external priors, we consider two different cases, representative of currently available data and of data likely to be available in the 2030s, when redshift drift measurements may be under way or possibly already available. Specifically, for current data we use the measurement of $\Omega_m h^2$ from the Planck 2018 \citep{Planck}, $\sigma_{\omega,Planck}=0.0013$ and uncertainties on the dark energy equation of state parameters $\sigma_{w0}=0.1$ and $\sigma_{wa}=0.3$, typical of Planck and DES \citep{Abbott}. For future data we rely on the recent detailed studies of the CORE collaboration \citep{CORE}, which forecasts $\sigma_{\omega,CORE}=0.00028$ and on the Euclid mission for which one expects $\sigma_{w0}=0.02$ and $\sigma_{wa}=0.1$ \citep{Euclid}.

Our diagnostics, to be listed in subsequent tables, are the correlation coefficients $\rho$ for the relevant pairs of parameters and the pairs' Figures of Merit (hereafter FoM, c.f. Eq. \ref{FoMs}) and the one-sigma marginalized uncertainties for each of the model parameters. In Appendix A we briefly discuss how the constraints are improved if one increases the integration times beyond the ones listed above.

\subsection{The $\Lambda$CDM case}

\begin{figure*}
\begin{center}
\includegraphics[width=3.2in,keepaspectratio]{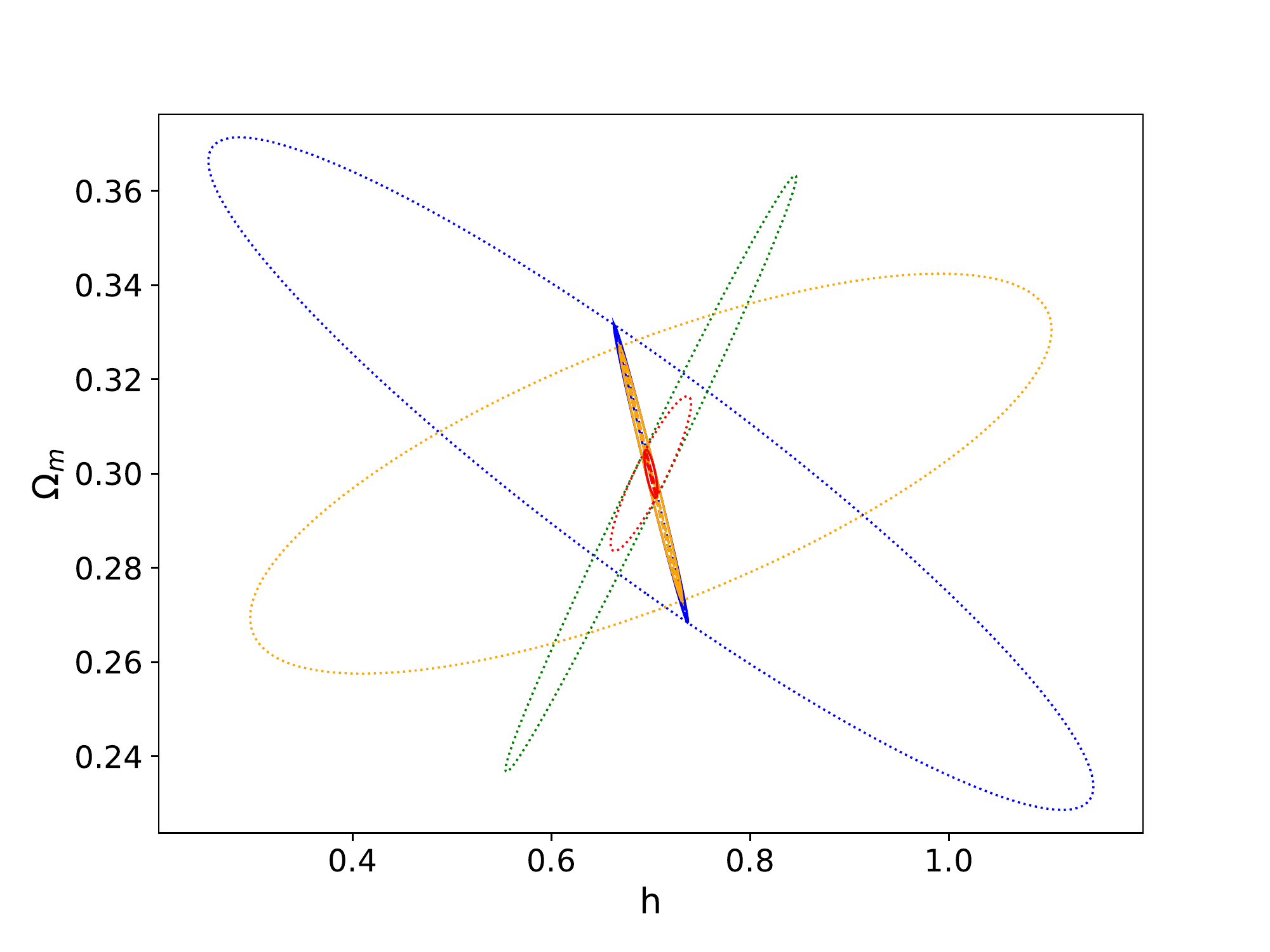}
\includegraphics[width=3.2in,keepaspectratio]{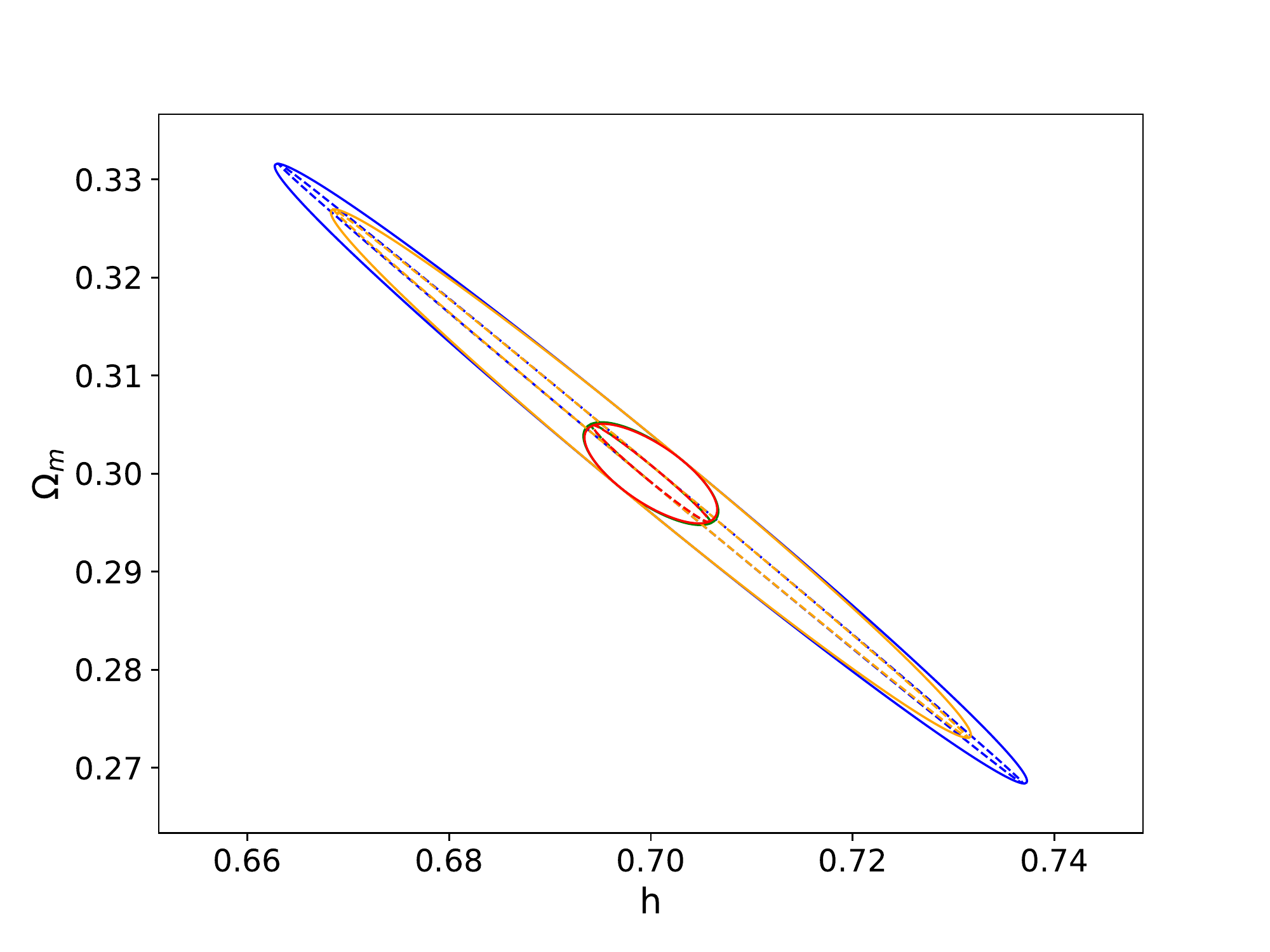}
\end{center}
\caption{One-sigma constraints on the $\Omega_m$--$h$ plane, for the $\Lambda$CDM model. The blue lines represent the ELT, the green ones the SKA, the yellow ones the CHIME and the red ones the combination of all three. The dotted lines are the constraints without priors, the solid lines the constraints with current priors and the dashed lines the constraints with future priors. The right panel is a close up of the left one.}
\label{fig7}
\end{figure*}

The results of this analysis are summarized in Table \ref{table1} and Fig. \ref{fig7}. Starting with measurements of each of the individual facilities, we note that both the ELT and the CHIME on their own can only get poor measurements of $h$, but reasonably competitive measurements of $\Omega_m$; for this reason they yield relatively small FoMs in the $\Omega_m$--$h$ plane. On the other hand the SKA is much more sensitive to $h$ while having a sensitivity on $\Omega_m$ which is intermediate between those of the ELT and CHIME. This leads to an overall FoM which is about 10 times larger than those of the ELT and CHIME. The combined measurements of all three facilities lead to one-sigma constraints $\sigma(\Omega_m)=0.011$ and $\sigma(h)=0.027$, with an overall FoM that is almost four times larger than that of SKA and 40 times larger than those of the ELT and CHIME.

The reason for the significant gains in the combination is the one we discussed in previous sections, namely that the degeneracy direction in the $\Omega_m$--$h$ plane rotates with redshift, with the correlation coefficient being negative for the ELT but positive for the SKA and CHIME. This is particularly clear in the left panel of Fig. \ref{fig7}. The addition of Planck or CORE priors of course leads to further improvements, with forecasted one-sigma constraints $\sigma(\Omega_m)=0.003$ and $\sigma(h)=0.004$; in this case the two parameters will always be anticorrelated, as a result of the prior itself.

\begin{table*}
\centering
\caption{Results of the Fisher Matrix analysis for the $\Lambda$CDM model. The first line shows the correlation coefficient $\rho$ for the two parameters, the next one the Figure of Merit (rounded to the nearest integer), and the last two the one-sigma marginalized uncertainties for the two parameters. The {\it All} case corresponds to the combination ELT$+$SKA$+$CHIME, while {\it C} and {\it F} respectively denote the current (Planck-like) and future (CORE-like) priors on $\Omega_mh^2$ discussed in the text.}
\label{table1}
\resizebox{\textwidth}{!}{%
\begin{tabular}{| c | c c c | c c c | c c c | c c c |}
\hline
Parameter & ELT & ELT+C & ELT+F & SKA & SKA+C & SKA+F & CHIME & CHIME+C & CHIME+F & All & All+C & All+F \\
\hline
$\rho(h,\Omega_m)$ & -0.930& -0.992& -0.9996&  0.993& -0.733 & -0.985 & 0.721 & -0.989 & -0.999 &  0.902 & -0.723 & -0.985 \\
\hline
$FoM(h,\Omega_m)$ & 86 & 6762 &  31391  & 903 &  42182  & 195803  & 85 &  7927   &  36802  &  3505   &  43582  & 201718 \\
\hline
$\sigma(\Omega_m)$ &0.047 & 0.021 &  0.021 &  0.042 &  0.003 &  0.003 &  0.028 &  0.018 &  0.018 &  0.011 &  0.003  &  0.003 \\
$\sigma(h)$ & 0.293  &  0.025  &  0.024  &  0.096  &  0.004  &  0.004  &  0.265  &  0.021  &  0.021  &  0.027  &  0.004  &  0.004  \\
\hline
\end{tabular}}
\end{table*}

\subsection{The $w_0$CDM case}

\begin{figure*}
\begin{center}
\includegraphics[width=3.2in,keepaspectratio]{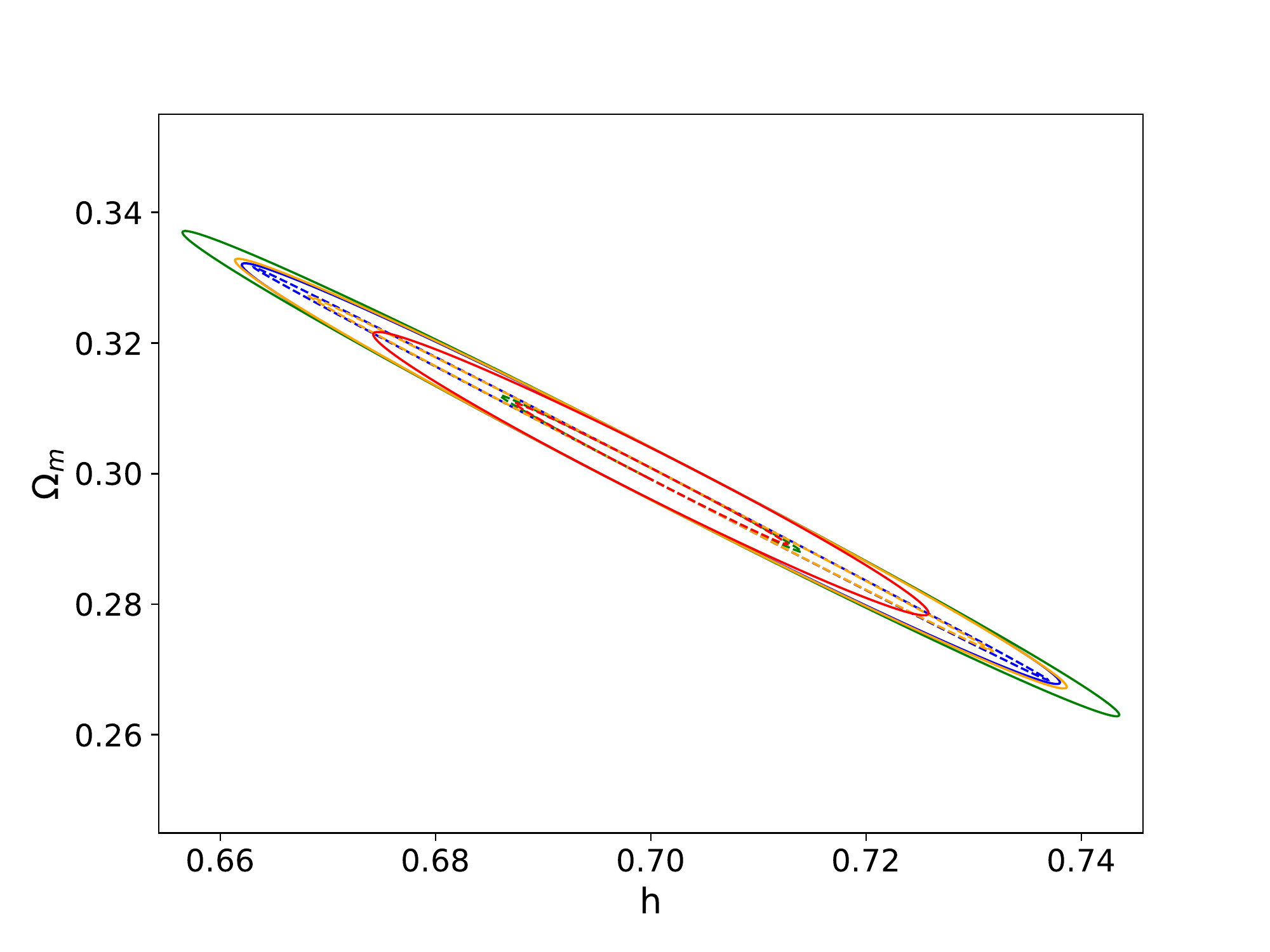}
\includegraphics[width=3.2in,keepaspectratio]{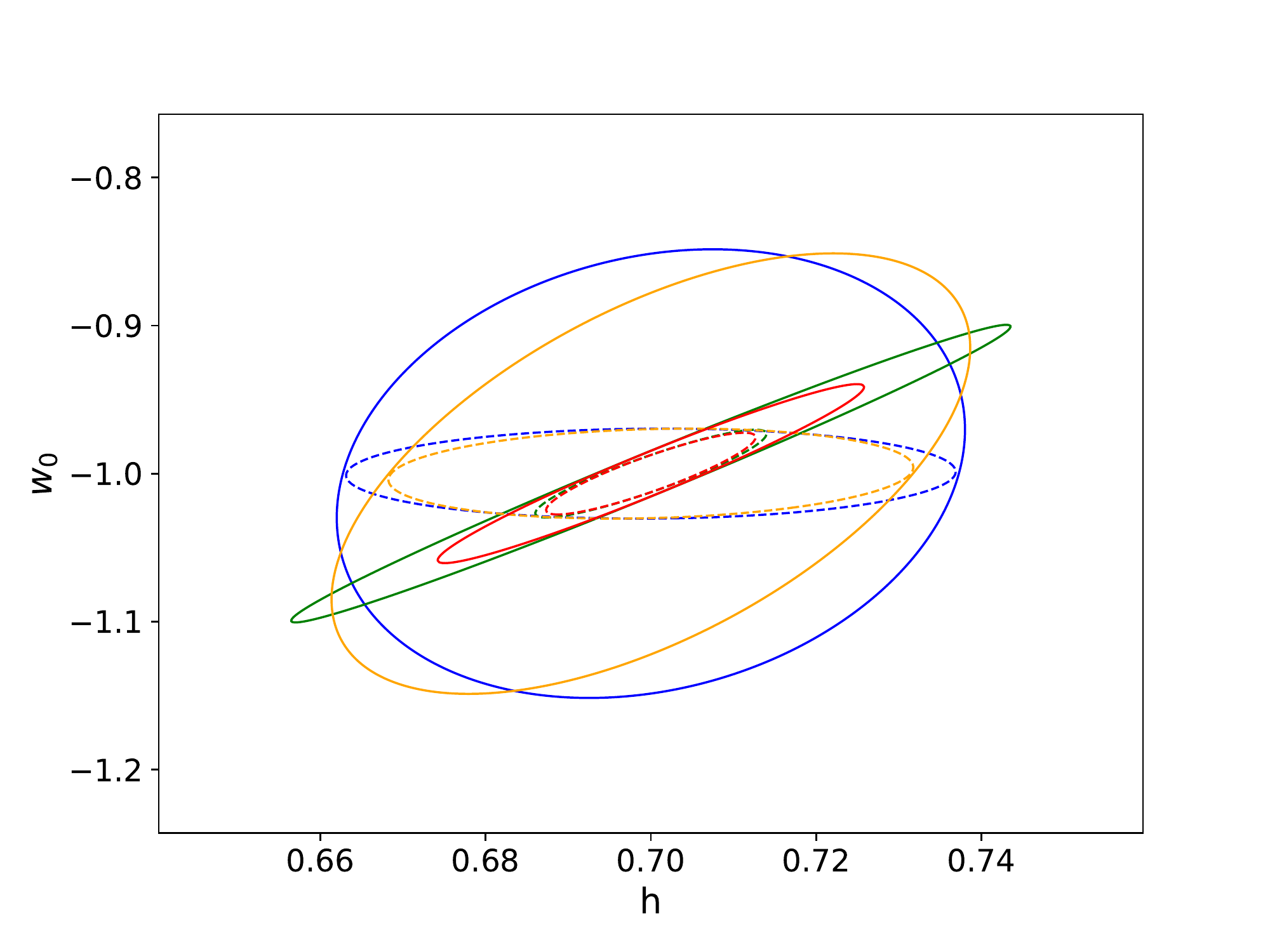}
\includegraphics[width=3.2in,keepaspectratio]{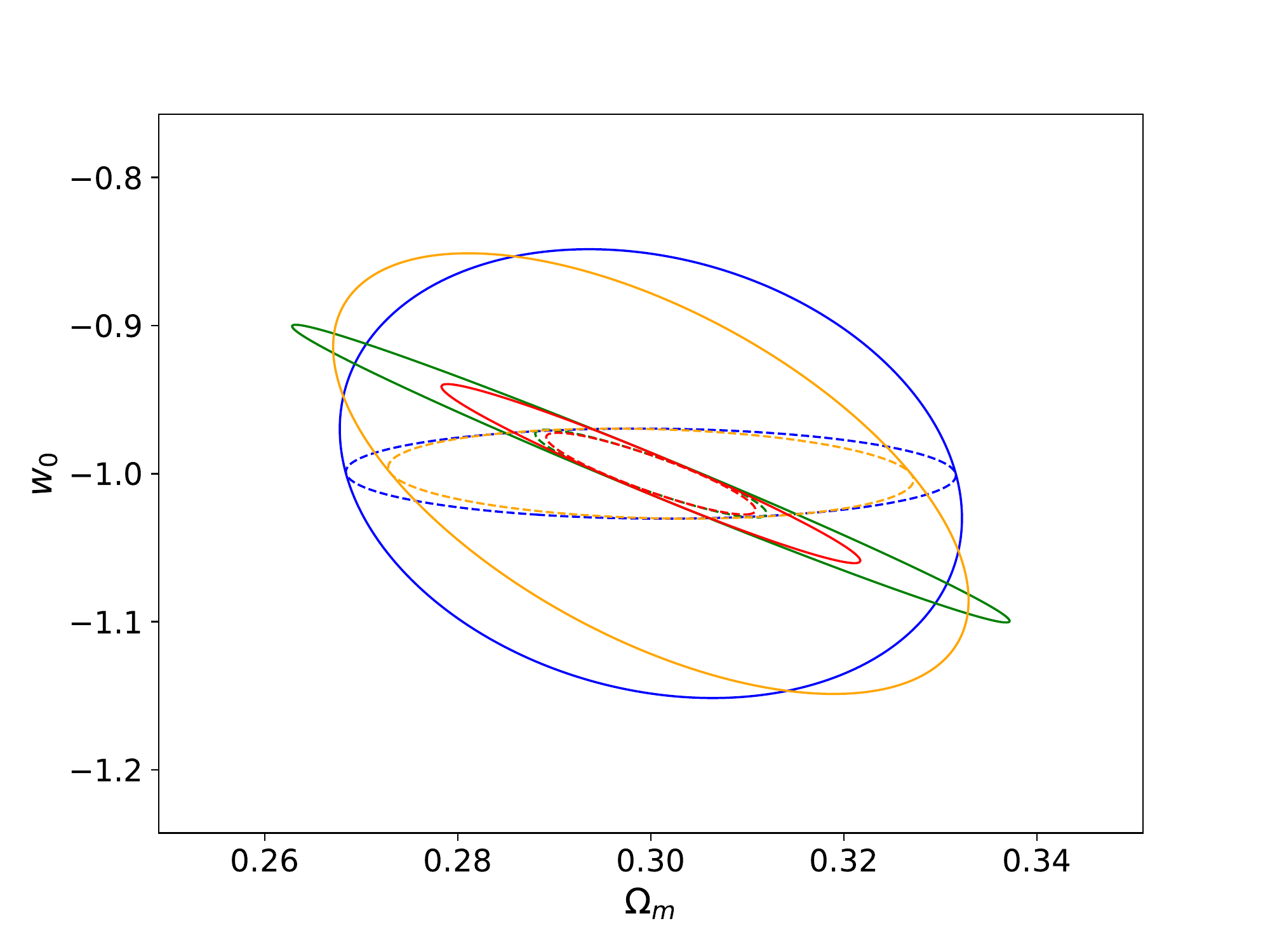}
\end{center}
\caption{One-sigma constraints in the three relevant 2D planes, for the $w_0$CDM model. The blue lines represent the ELT, the green ones the SKA, the yellow ones the CHIME and the red ones the combination of all three. The solid lines show the constraints with current priors and the dashed lines the constraints with future priors.}
\label{fig8}
\end{figure*}

The results of this analysis are summarized in Table \ref{table2} and Fig. \ref{fig8}, which can be compared to those in the previous sub-section. In this case the extension of parameter space with a constant equation of state, $w_0$, which need not be a cosmological constant, means that constraints from the three individual facilities (without the addition of external priors) are weaker. Nevertheless the combination of the three facilities still leads to very significant constraints, specifically $\sigma(\Omega_m)=0.02$ for the matter density and $\sigma(w_0)=0.13$ for the dark energy equation of state. On the other hand, for the Hubble constant we get the weaker constraint $\sigma(h)=0.13$.

The addition of priors on $\Omega_mh^2$ enables the ELT or CHIME to constrain the matter density to a level comparable to what can be obtained for all three facilities combined, with the ELT leading to the best constraint (a natural consequence of probing higher redshifts). The priors also lead to a significantly improvement on the constraint on the Hubble parameter, and the combination of all three facilities with future priors will constrain $\sigma(h)=0.008$.

As for the dark energy equation of state $w_0$, the ELT or CHIME can do no better than the priors, while the SKA can improve on the constraint from current priors though not on that from future priors. The three facilities together with current priors can constrain $\sigma(w_0)=0.04$ (which is a $60\%$ improvement on the $w_0$ prior itself) while with future priors the constraint is $\sigma(w_0)=0.018$ (which is still a $10\%$ improvement on the prior).

\begin{table*}
\centering
\caption{Results of the Fisher Matrix analysis for the $w_0$CDM model. The first set of lines show the correlation coefficients $\rho$ for each pair of parameters, the following ones the Figure of Merit for each pair of parameters (rounded to the nearest integer), and the next ones the one-sigma marginalized uncertainties for each of the parameters. The final row shows how much the redshift drift measurement improves on the $w_0$ prior. The {\it All} case corresponds to the combination ELT$+$SKA$+$CHIME, while {\it C} and {\it F} respectively denote the current (Planck-like) and future (CORE-like and Euclid-like) priors on $\Omega_mh^2$ and $w_0$ discussed in the text.}
\label{table2}
\resizebox{\textwidth}{!}{%
\begin{tabular}{| c | c c c | c c c | c c c | c c c |}
\hline
Parameter & ELT & ELT+C & ELT+F & SKA & SKA+C & SKA+F & CHIME & CHIME+C & CHIME+F & All & All+C & All+F \\
\hline
$\rho(h,\Omega_m)$ & -0.996 & -0.992  & -1.000  & -0.991  & -0.994  & -0.997  & -0.980 & -0.993  & -0.999  & -0.714  & -0.983  & -0.997 \\
$\rho(\Omega_m,w_0)$& -0.994 & -0.199 & -0.041 & 0.988  & -0.990 & -0.906 & -0.995 & -0.575 & -0.142 & -0.833 & -0.972 & -0.891 \\
$\rho(h,w_0)$ & 0.984  & 0.197  & 0.041  & 1.000  & 0.988  & 0.905  & 0.993  & 0.571  & 0.142  & 0.979  & 0.967  & 0.890 \\
\hline
$FoM(h,\Omega_m)$ & 6.802  &  6627  & 31365  & 3.196  &  5745  & 82725  & 3.438  &  6487  & 36429  &  244   &  9846  & 91388 \\
$FoM(\Omega_m,w_0)$& 1.596  &  210   &  1046  & 4.089  &  1907  &  6682  & 4.726  &  251   &  1228  &  311   &  3275  &  7396 \\
$FoM(h,w_0)$ & 0.256  &  178   &  897   & 1.768  &  1495  &  5700  & 0.497  &  213   &  1052  &  126   &  2527  &  6303 \\
\hline
$\sigma(\Omega_m)$ & 0.436  & 0.021  & 0.021  & 0.269  & 0.024  & 0.008  & 0.273  & 0.022  & 0.018  & 0.020  & 0.014  & 0.007\\
$\sigma(h)$ & 1.620  & 0.025  & 0.024  & 3.880  & 0.029  & 0.009  & 2.311  & 0.025  & 0.021  & 0.130  & 0.017  & 0.008  \\
$\sigma(w_0)$ & 5.806  & 0.100  & 0.020  & 2.555  & 0.066  & 0.020  & 3.296  & 0.098  & 0.020  & 0.130  & 0.040  & 0.018 \\
\hline
$Gain(w_0)$&   -    & $<1\%$  & $<1\%$  &   -    & $34\%$  & $3\%$  &   -    & $2\%$  & $<1\%$  &   -    & $60\%$  & $9\%$ \\
\hline
\end{tabular}}
\end{table*}

\subsection{The CPL case}

\begin{figure*}
\begin{center}
\includegraphics[width=3.2in,keepaspectratio]{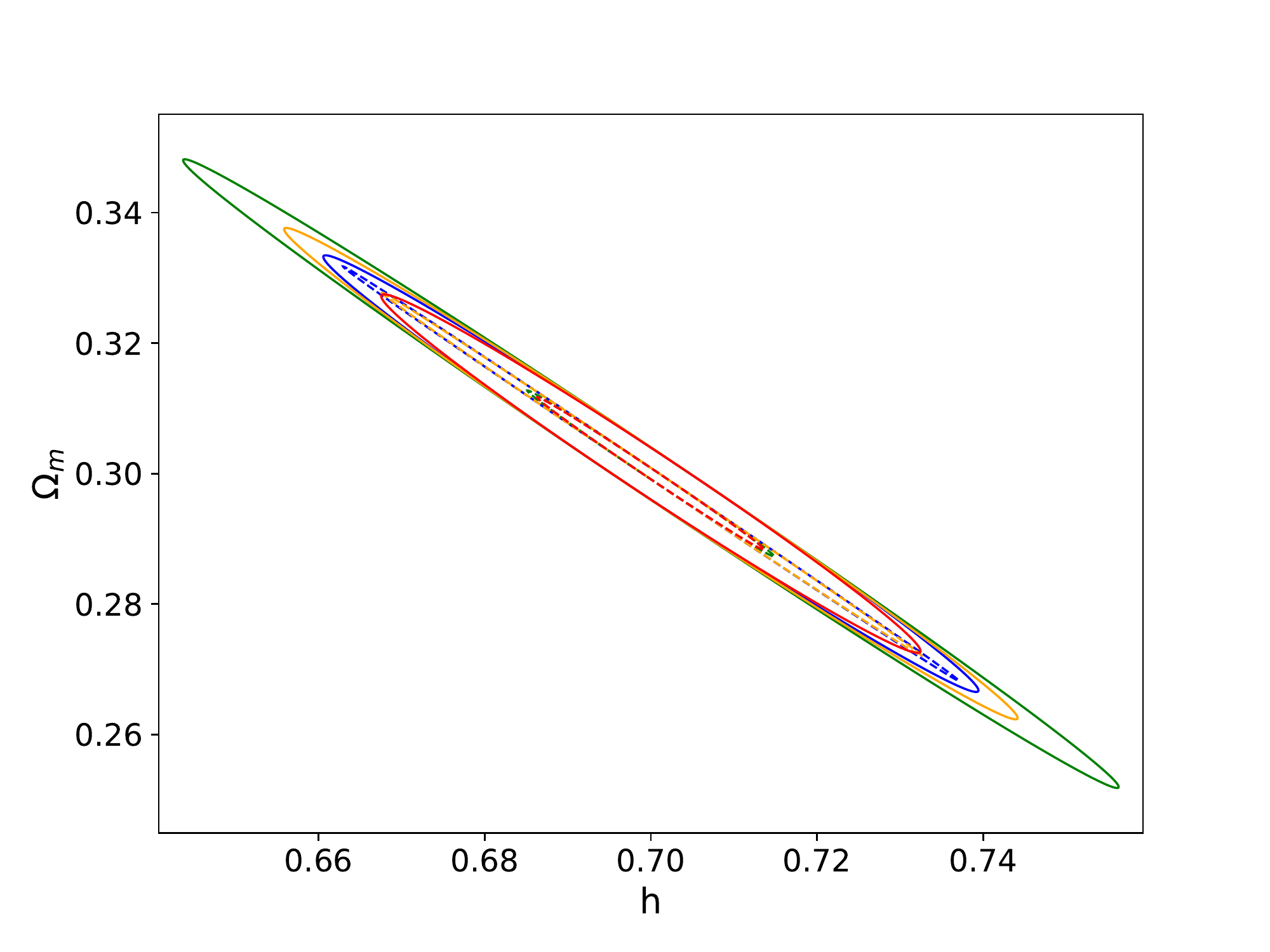}
\includegraphics[width=3.2in,keepaspectratio]{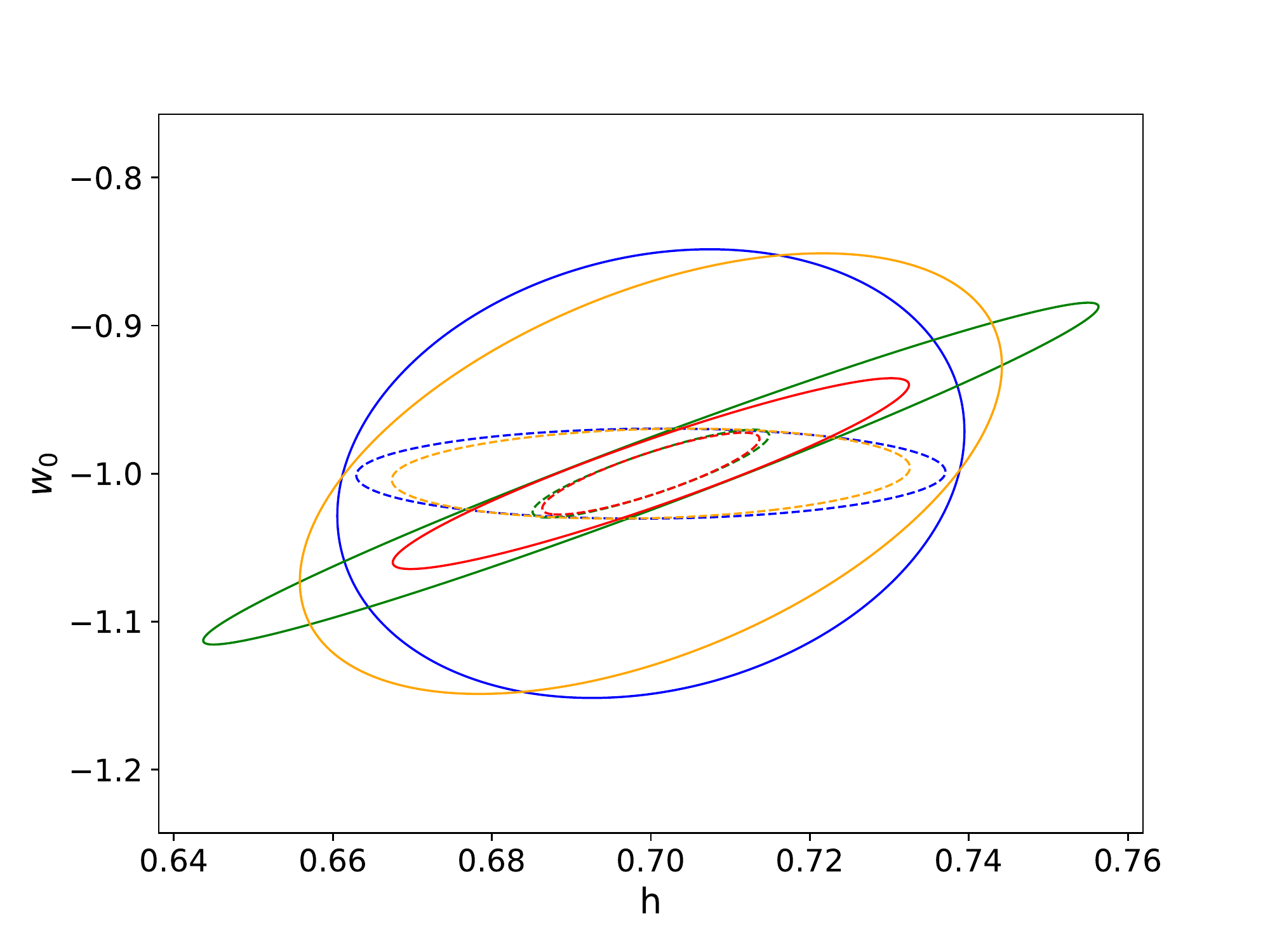}
\includegraphics[width=3.2in,keepaspectratio]{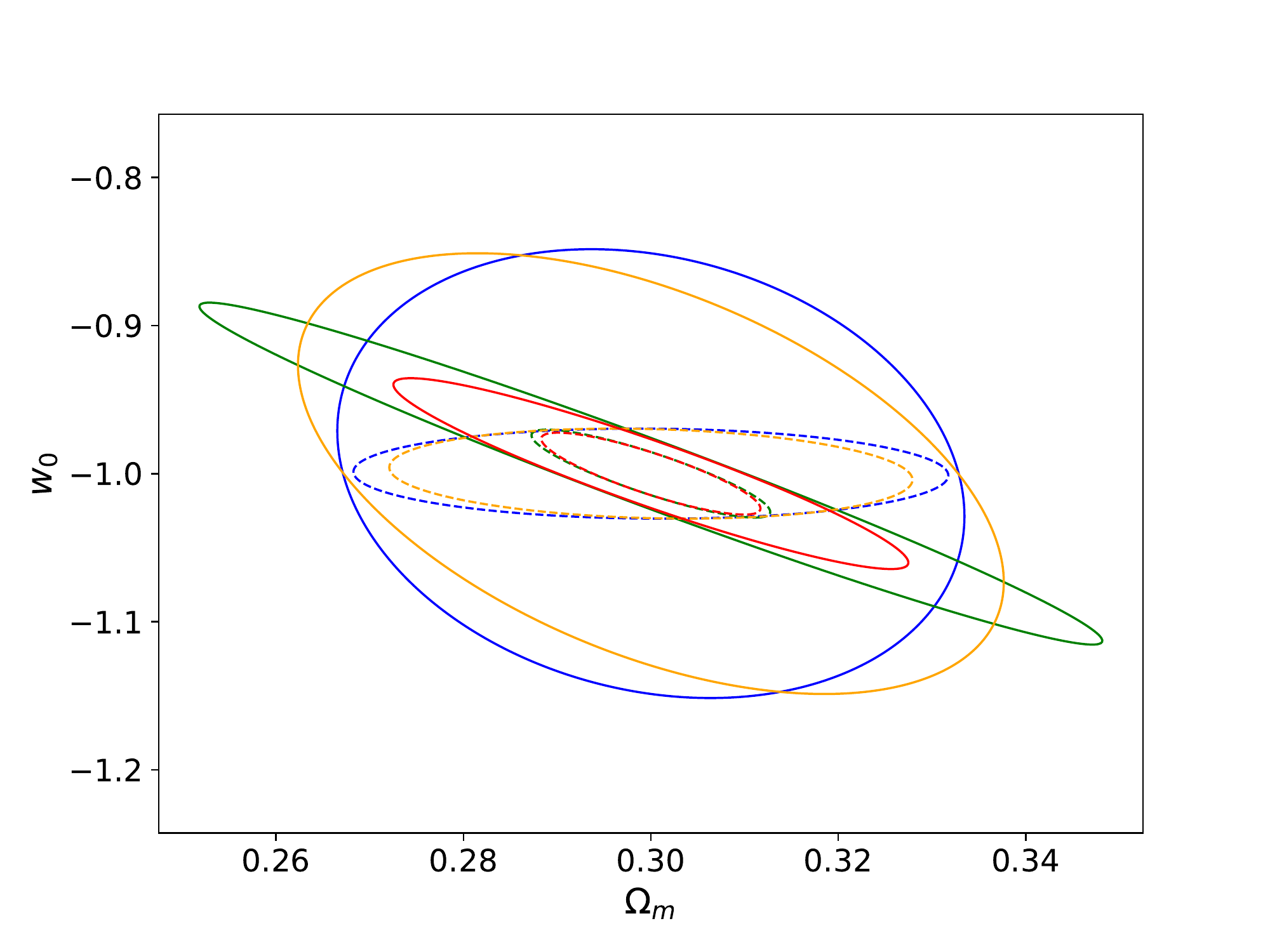}
\includegraphics[width=3.2in,keepaspectratio]{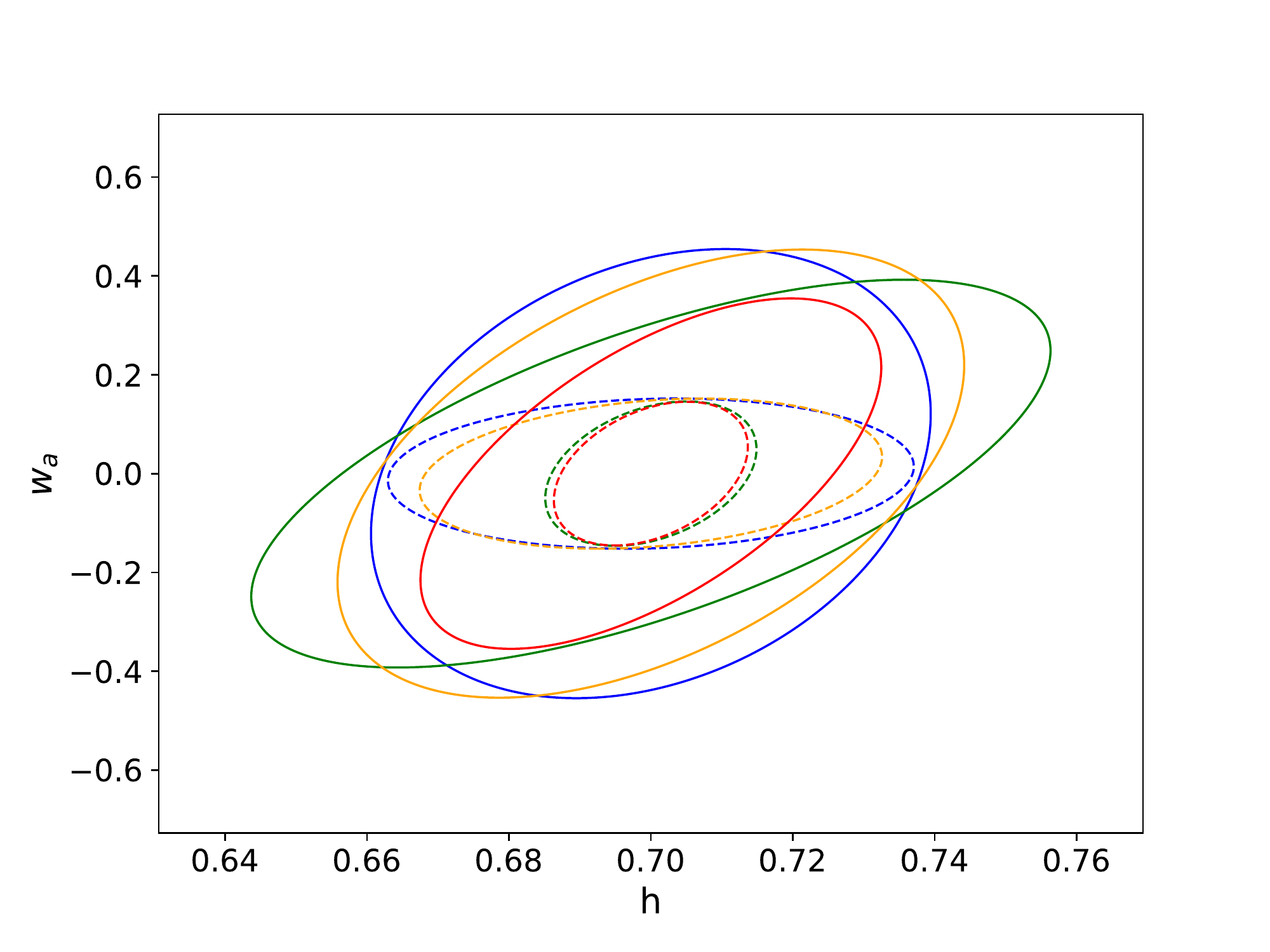}
\includegraphics[width=3.2in,keepaspectratio]{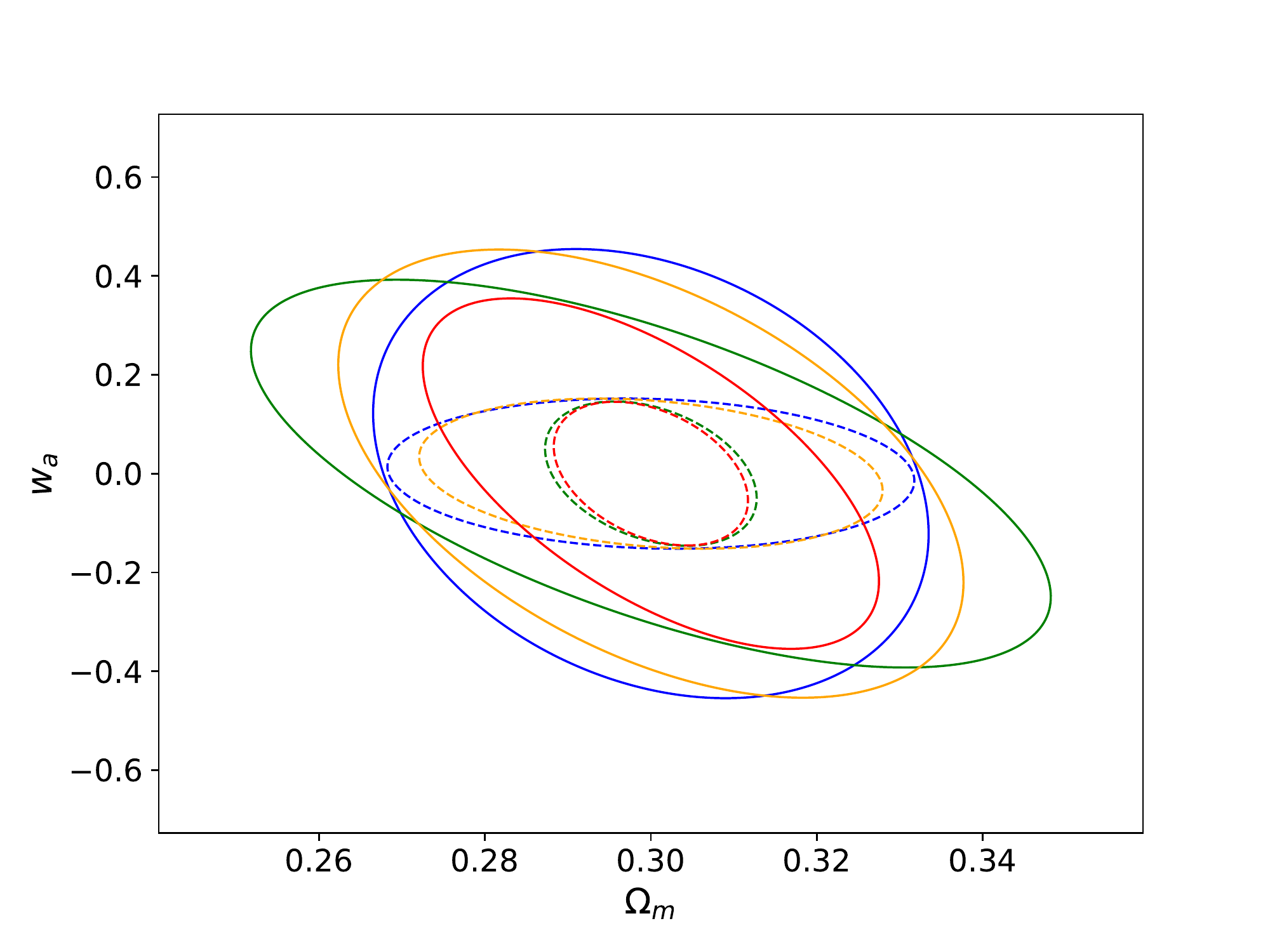}
\includegraphics[width=3.2in,keepaspectratio]{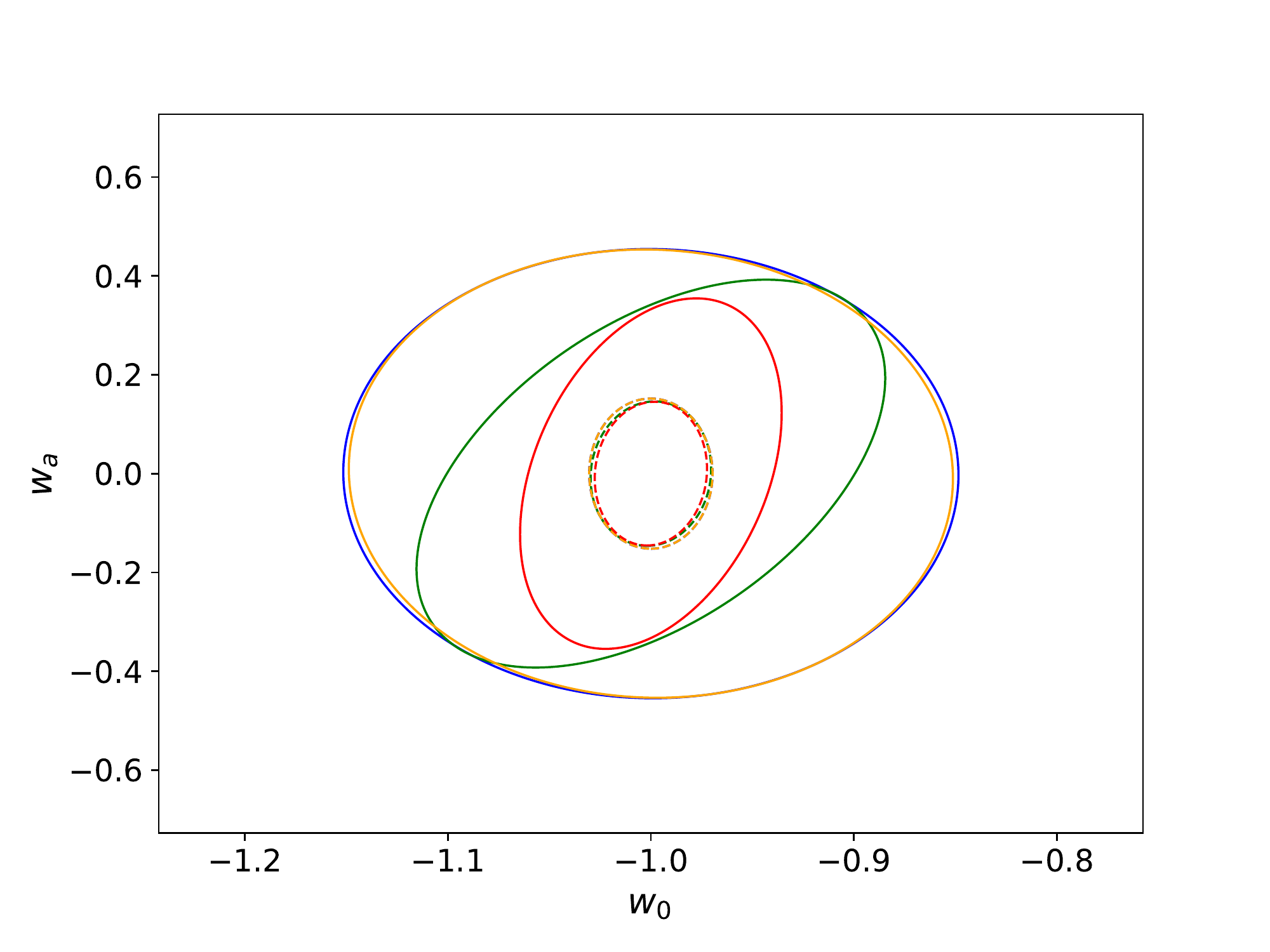}
\end{center}
\caption{One-sigma constraints in the six relevant 2D planes, for the CPL model. The blue lines represent the ELT, the green ones the SKA, the yellow ones the CHIME and the red ones the combination of all three. The solid lines show the constraints with current priors and the dashed lines the constraints with future priors.}
\label{fig9}
\end{figure*}

The results of this analysis are summarized in Table \ref{table3} and Fig. \ref{fig9}. The extension of the parameter space with an additional parameter (in this case describing the evolution of the dark energy equation of state) naturally leads to weaker constraints. Nevertheless, a direct mapping of the expansion of the universe from $z=0$ to beyond $z=4$, without any priors, can constrain the matter density to $\sigma(\Omega_m)=0.068$ and the dark energy equation of state to $\sigma(w_0)=0.320$ and $\sigma(w_a)=0.822$. Again the Hubble parameter is the less well constrained parameter, in this case with the rather large $\sigma(h)=0.322$.

The addition of priors naturally leads to improved constraints. Focusing on the dark energy equation of state parameters, we note that the ELT or CHIME can't improve on the $w_0$ or $w_a$ priors, while the SKA can improve the constraints from current priors by about $24\%$ and $14\%$ respectively. However, combined measurements from the three facilities with current priors can improve on them by $58\%$ and $22\%$ respectively. Relative to future Euclid-like priors the improvements are more modest, at the $10\%$ and $4\%$ level respectively.

For most cosmological probes, the dark energy parameters $w_0$ and $w_a$ will be anticorrelated. However, this need not be the case for the redshift drift, as has been briefly pointed out in \citet{Kim}. Indeed, for the redshift drift the said behaviour will be redshift-dependent: the correlation will be positive at low redshifts and negative at high redshifts. As Table \ref{table3} shows, for the combined measurements of the three facilities, without any priors, there is a strong positive correlation (since the combination is dominated by the SKA measurements), which is decreased by the addition of priors. However, when one looks at the individual facilities (with priors), the two parameters are almost entirely uncorrelated for the cases of the ELT and CHIME, and mildly positively correlated for the case of the SKA.

In the context of forecasts for the CPL model one often discusses the pivot redshift \citep{FMA1,Huterer}, defined as
\be
z_p=\frac{-1}{1+\frac{\sigma(w_a)}{\rho\sigma(w_0)}}\,.
\ee
This is usually interpreted as the redshift at which the dark energy is best constrained. We note that this interpretation assumes (at least implicitly) that the redshift will be positive, which will only be the case if the two dark energy parameters are anticorrelated. For the redshift drift, since this need not be the case, one may have negative pivot redshifts, as shown in the last row of Table \ref{table3}. This issue of the sign somewhat calls into question the physical meaning (and usefulness) of the pivot redshift concept for generic cosmological observables. Still, and despite this caveat, it is also worthy of note that in almost cases the pivot redshift is effectively zero, which is commensurate with the fact that when doing measurements of the redshift drift as discussed in this work we are taking the present days as the comparison point.

\begin{table*}
\centering
\caption{Results of the Fisher Matrix analysis for the CPL model. The first set of lines show the correlation coefficients $\rho$ for each pair of parameters, the next set the Figure of Merit for each pair of parameters (rounded to the nearest integer), and next ones the one-sigma marginalized uncertainties for each of the parameters, together with how much the redshift drift measurement improves on the $w_0$ and $w_a$ priors. The last line shows the pivot redshifts discussed in the main text. The {\it All} case corresponds to the combination ELT$+$SKA$+$CHIME, while {\it C} and {\it F} respectively denote the current (Planck-like) and future (CORE-like and Euclid-like) priors on $\Omega_mh^2$, $w_0$ and $w_a$ discussed in the text.}
\label{table3}
\resizebox{\textwidth}{!}{%
\begin{tabular}{| c | c c c | c c c | c c c | c c c |}
\hline
Parameter & ELT & ELT+C & ELT+F & SKA & SKA+C & SKA+F & CHIME & CHIME+C & CHIME+F & All & All+C & All+F \\
\hline
$\rho(h,\Omega_m)$& -1.000 & -0.993 & -1.000 & -0.570 & -0.997 & -0.998 & -0.999 & -0.994 & -1.000 & -0.959 & -0.989 & -0.997  \\
$\rho(\Omega_m,w_0)$& 0.983  & -0.190 & -0.040 & -0.578 & -0.978 & -0.871 & -0.848 & -0.492 & -0.138 & -0.973 & -0.933 & -0.850\\
$\rho(h,w_0)$& -0.978 & 0.188  & 0.040  & 1.000  & 0.977  & 0.871  & 0.872  & 0.489  & 0.138  & 0.997  & 0.931  & 0.850 \\
$\rho(\Omega_m,w_a)$& -0.995 & -0.269 & -0.095 & -0.636 & -0.636 & -0.341 & -0.998 & -0.486 & -0.224 & -0.958 & -0.614 & -0.374 \\
$\rho(h,w_a)$& 0.992  & 0.266  & 0.095  & 0.997  & 0.635  & 0.340  & 0.995  & 0.484  & 0.224  & 0.914  & 0.607  & 0.374 \\
$\rho(w_0,w_a)$ & -0.996 & -0.007 & -0.000 & 0.997  & 0.492  & 0.062  & 0.816  & -0.022 & -0.001 & 0.914  & 0.346  & 0.069\\
\hline
$FoM(h,\Omega_m)$& 0.244  &  6382  & 31223  & 0.032  &  4432  & 77777  & 0.083  &  5669  & 35501  &   70   &  7775  & 84737\\
$FoM(\Omega_m,w_0)$& 0.008  &  202   &  1042  & 0.044  &  864   &  5407  & 0.030  &  206   &  1196  &   86   &  1574  &  5897\\
$FoM(h,w_0)$ & 0.002  &  171   &  892   & 0.025  &  725   &  4620  & 0.007  &  175   &  1025  &   51   &  1316  &  5036 \\
$FoM(\Omega_m,w_a)$& 0.003  &   69   &  209   & 0.061  &   69   &  574   & 0.012  &   67   &  243   &   27   &  131   &  637\\
$FoM(h,w_a)$ & 0.001  &   58   &  179   & 0.004  &   59   &  491   & 0.001  &   57   &  208   &   4    &  110   &  545 \\
$FoM(w_0,w_a)$& 0.000  &   15   &  218   & 0.006  &   25   &  232   & 0.001  &   15   &  218   &   4    &   47   &  250\\
\hline
$\sigma(\Omega_m)$ & 4.568  & 0.022  & 0.021  & 0.348  & 0.032  & 0.008  & 4.772  & 0.025  & 0.018  & 0.068  & 0.018  & 0.008\\
$\sigma(h)$& 13.129 & 0.026  & 0.024  & 46.788 & 0.037  & 0.010  & 22.140 & 0.029  & 0.021  & 0.322  & 0.021  & 0.009\\
$\sigma(w_0)$& 66.768 & 0.100  & 0.020  & 35.186 & 0.076  & 0.020  & 5.707  & 0.098  & 0.020  & 0.320  & 0.042  & 0.018\\
$\sigma(w_a)$& 302.459 & 0.299  & 0.100  & 26.595 & 0.258  & 0.096  & 130.763 & 0.298  & 0.100  & 0.822  & 0.233  & 0.096\\
\hline
$Gain(w_0)$&   -    & $<1\%$  & $<1\%$  &   -    & $24\%$  & $2\%$  &   -    & $2\%$  & $<1\%$  &   -    & $58\%$  & $9\%$ \\
$Gain(w_a)$&   -    & $<1\%$  & $<1\%$  &   -    & $14\%$  & $4\%$  &   -    & $<1\%$  & $<1\%$  &   -    & $22\%$  & $4\%$ \\
\hline
$z_p$& 0.282  & 0.002  & 0.000  & -0.569 & -0.126 & -0.012 & -0.034 & 0.007  & 0.000  & -0.263 & -0.059 & -0.013 \\
\hline
\end{tabular}}
\end{table*}

\subsection{On the choice of fiducial model}

\begin{figure*}
\begin{center}
\includegraphics[width=3.2in,keepaspectratio]{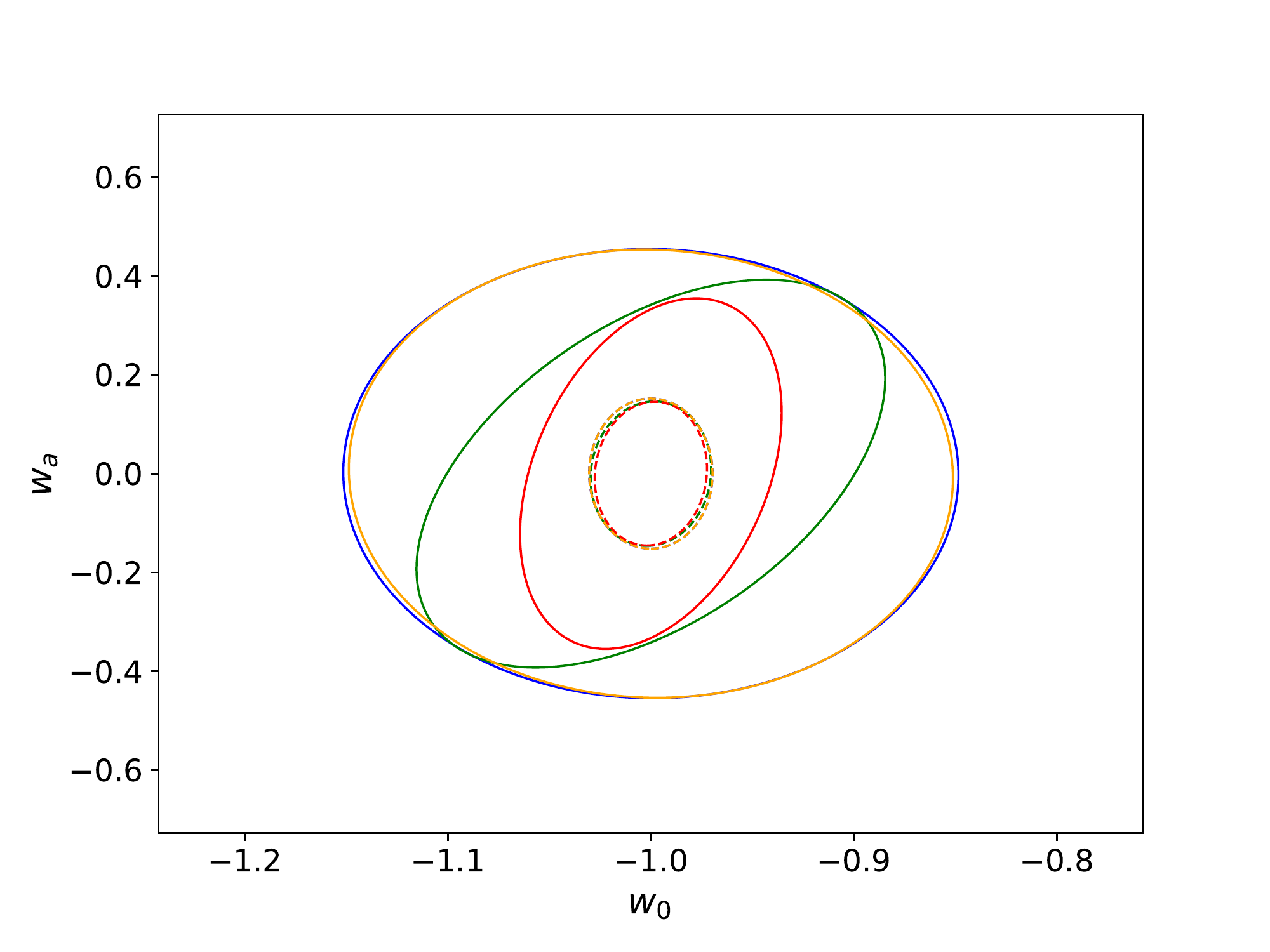}
\vskip0in
\includegraphics[width=3.2in,keepaspectratio]{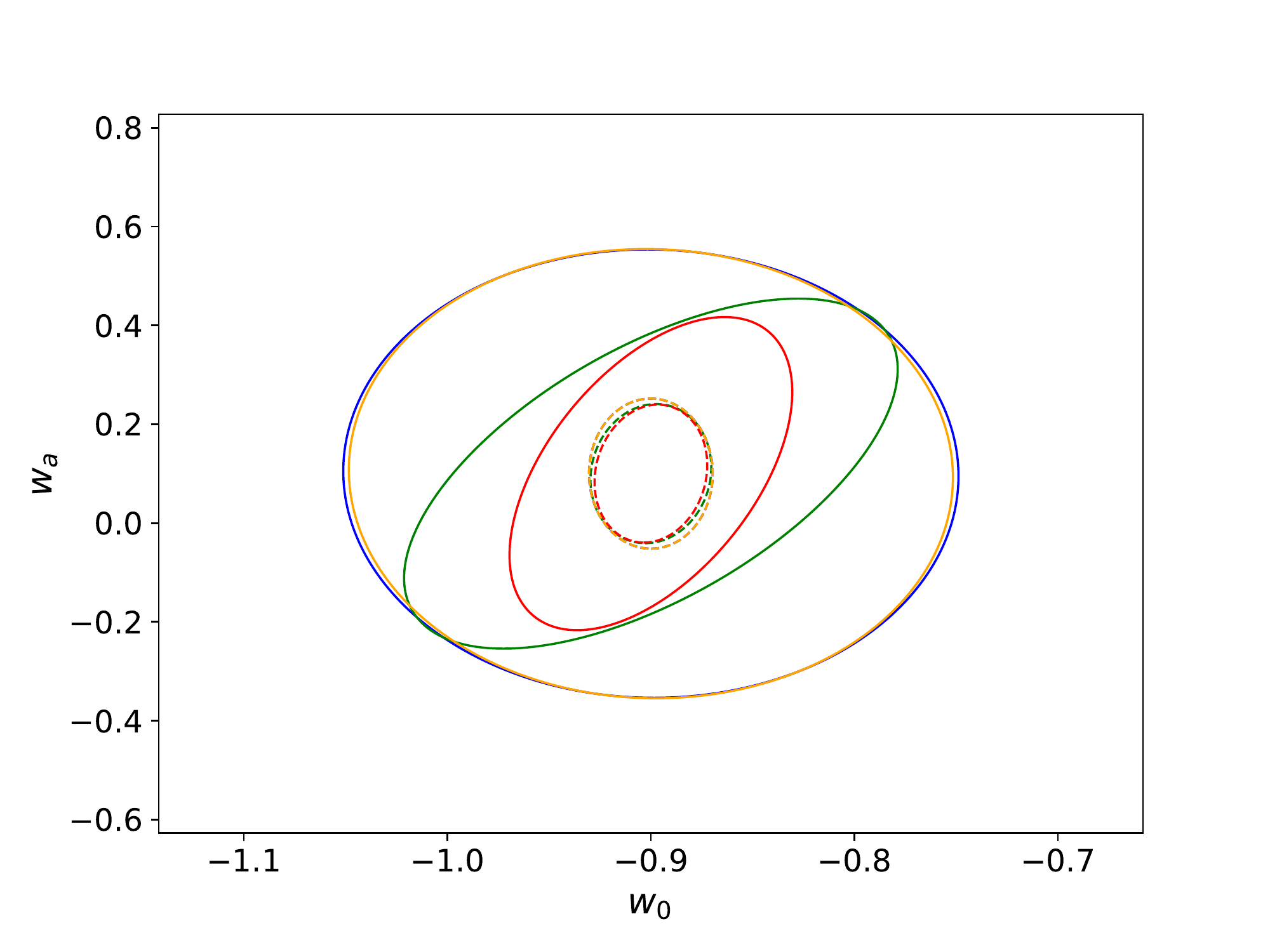}
\includegraphics[width=3.2in,keepaspectratio]{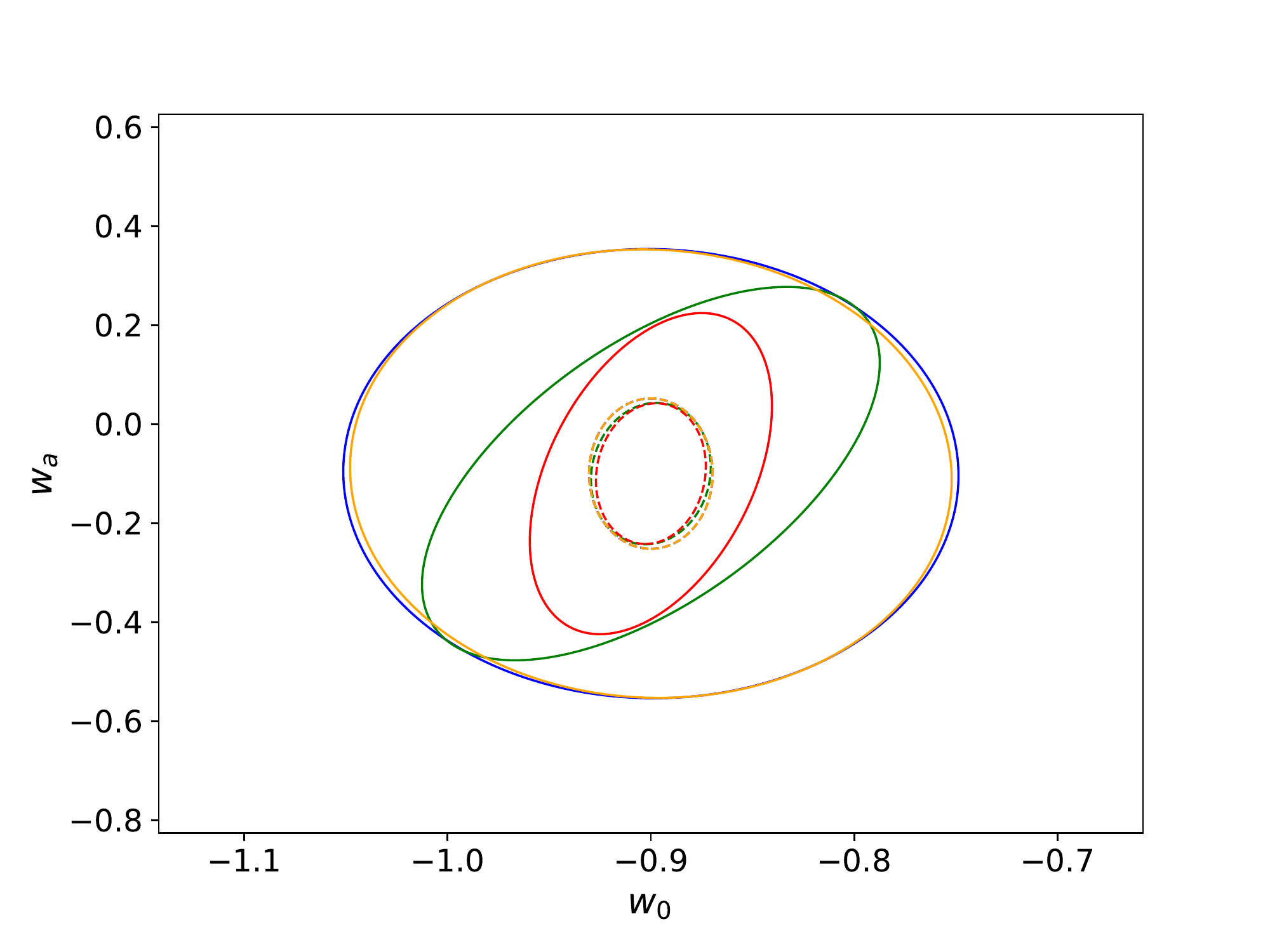}
\end{center}
\caption{One-sigma constraints in the $w_0$-$w_a$ plane, for three choices of CPL fiducial model: the top panel corresponds to $\Lambda$CDM, while the bottom left and right panels respectively correspond to the freezing and thawing models defined in the text. The blue lines represent the ELT, the green ones the SKA, the yellow ones the CHIME and the red ones the combination of all three. The solid lines show the constraints with current priors and the dashed lines the constraints with future priors.}
\label{fig10}
\end{figure*}

Finally, we briefly discuss the dependence of our results on the choice of fiducial model, specifically for the dark energy sector. In the previous sub-section our fiducial model was standard $\Lambda$CDM ($w_0=-1, w_a=0$). Here we will compare the results for this and two other fiducial models, ($w_0=-0.9, w_a=+0.1$) and ($w_0=-0.9, w_a=-0.1$), which are examples of freezing and thawing models in the phenomenological classification of \citet{Caldwell}. The results are summarized in Table \ref{table4} and Fig. \ref{fig10}.

\begin{table*}
\centering
\caption{Comparison of the results of the Fisher Matrix analysis for three possible choices of fiducial parameters in the CPL model. The top third of the table repeats the dark energy related part of Table \ref{table3}, which corresponds to a $\Lambda$CDM fiducial model. The middle and bottom thirds of the table show analogous results for freezing and thawing fiducial models. The parameters and combinations of datasets are as in the previous tables.}
\label{table4}
\resizebox{\textwidth}{!}{%
\begin{tabular}{| c | c | c c c | c c c | c c c | c c c |}
\hline
Fiducial & Parameter & ELT & ELT+C & ELT+F & SKA & SKA+C & SKA+F & CHIME & CHIME+C & CHIME+F & All & All+C & All+F \\
\hline
{ } & $\rho(w_0,w_a)$& -0.996 & -0.007 & -0.000 & 0.997  & 0.492  & 0.062  & 0.816  & -0.017 & -0.001 & 0.914  & 0.346  & 0.069\\
$w_0=-1.0$ & $FoM(w_0,w_a)$ & 0.000  &   15   &  218   & 0.006  &   25   &  232   & 0.001  &   15   &  218   &   4    &   47   &  250\\
{ } & $\sigma(w_0)$ & 66.768 & 0.100  & 0.020  & 35.186 & 0.076  & 0.020  & 5.707  & 0.098  & 0.020  & 0.320  & 0.042  & 0.018\\
$w_a=0.0$ & $\sigma(w_a)$& 302.459 & 0.299  & 0.100  & 26.595 & 0.258  & 0.096  & 130.763 & 0.299  & 0.100  & 0.822  & 0.233  & 0.096\\
{ } & $z_p$& 0.282  & 0.002  & 0.000  & -0.569 & -0.126 & -0.012 & -0.034 & 0.006  & 0.000  & -0.263 & -0.059 & -0.013 \\
\hline
{ } & $\rho(w_0,w_a)$& -0.988 & -0.011 & -0.001 & 0.997  & 0.597  & 0.090  & 0.982  & -0.017 & -0.001 & 0.943  & 0.521  & 0.124\\
$w_0=-0.9$ & $FoM(w_0,w_a)$ & 0.000  &   15   &  218   & 0.006  &   29   &  241   & 0.001  &   15   &  218   &   4    &   53   &  262 \\
{ } & $\sigma(w_0)$ & 34.677 & 0.099  & 0.020  & 41.858 & 0.080  & 0.020  & 20.746 & 0.098  & 0.020  & 0.435  & 0.046  & 0.018\\
$w_a=+0.1$ & $\sigma(w_a)$ & 245.152 & 0.298  & 0.100  & 23.423 & 0.233  & 0.093  & 123.303 & 0.299  & 0.100  & 0.756  & 0.208  & 0.092\\
{ } & $z_p$ & 0.162  & 0.004  & 0.000  & -0.641 & -0.170 & -0.019 & -0.142 & 0.006  & 0.000  & -0.352 & -0.103 & -0.024\\
\hline
{ } & $\rho(w_0,w_a)$ & -0.996 & -0.009 & -0.001 & 0.998  & 0.594  & 0.090  & 0.901  & -0.027 & -0.002 & 0.926  & 0.421  & 0.101\\
$w_0=-0.9$ & $FoM(w_0,w_a)$& 0.000  &   15   &  218   & 0.009  &   29   &  239   & 0.001  &   15   &  218   &   5    &   57   &  263\\
{ } & $\sigma(w_0)$& 52.791 & 0.100  & 0.020  & 32.605 & 0.074  & 0.019  & 7.006  & 0.097  & 0.020  & 0.298  & 0.039  & 0.018\\
$w_a=-0.1$ & $\sigma(w_a)$& 248.204 & 0.299  & 0.100  & 25.691 & 0.248  & 0.094  & 111.735 & 0.298  & 0.100  & 0.706  & 0.213  & 0.093 \\
{ } & $z_p$ & 0.269  & 0.003  & 0.000  & -0.559 & -0.151 & -0.018 & -0.053 & 0.009  & 0.000  & -0.281 & -0.072 & -0.019\\
\hline
\end{tabular}}
\end{table*}

One can see that the differences between the three cases are quite small. Nevertheless it is worth pointing out that when combining the three probes without external priors the best constraints, both for $w_0$ and for $w_a$, are obtained for the thawing model (which is again mainly due to the low-redshift sensitivity of the SKA). On the other hand, the weakest constraint on $w_0$ occurs in the freezing model, while the weakest constraint on $w_a$ occurs for $\Lambda$CDM.

\section{Conclusions}

We have presented an analysis of the cosmological impact of forthcoming redshift drift measurements, specifically from the ELT \citep{HIRES}, SKA \citep{Klockner} and CHIME \citep{Chime}, on constraints on cosmological model parameters. As previously mentioned, our CHIME results should also be broadly applicable to HIRAX \citep{HIRAX}. Our analysis used standard Fisher Matrix techniques, and we considered redshift drift measurements of each facility on its own and in combination, and also using additional priors representative of current and future observations. For concreteness we used three common fiducial models with different assumptions on the dark energy sector ($\Lambda$CDM, $w_0$CDM and CPL), but our formalism is generally applicable, as is the model-independent part of our analysis in the first half of this work.

In the literature there are already some forecasts of the cosmological impact of redshift drift measurements by the ELT, either using the specs from \cite{Liske} or some more optimistic version thereof. Our main contribution was to do a similar analysis for the SKA and CHIME, and thus also to compare the three for this purpose and to discuss synergies between them. Our goal has been to forecast the cosmological impact of redshift drift measurements by the three previously mentioned facilities, and not to forecast the sensitivity of the redshift measurements of each of these facilities---for the latter we have relied on the available published literature.

We must emphasize that conceptually a measurement of the redshift drift is of fundamental importance, as it will be the first time in our exploration of the distant universe that we will compare different past light cones---or, in other words, that we will see the universe expand in real time. Operationally, it is therefore a new probe of the universe---different from the ones we currently use. We believe that this alone is reason enough for these measurements to be done. Related to this point, the redshift drift is a direct and (in principle) model-independent probe of the expansion of the universe, and in particular of its acceleration phase. In practice, its role will likely be of a consistency test: standard cosmological observables will lead---under some modelling assumptions---to a predicted expansion history of the universe, $E(z)$, and such a history will imply a prediction of the value of the redshift drift as a function of redshift. A direct measurement of the signal at some observationally convenient redshift will then support or rule out these modelling assumptions.

Also for practical reasons our analysis focused on a more easily quantifiable aspect of the redshift drift measurements: their role in constraining dark energy scenarios. We confirm earlier suggestions that although on their own they yield comparatively weak constraints (at least if one allows for generic models with many cosmological parameters), they do probe regions of parameter space that are typically different from those probed by other experiments, as well as being redshift-dependent. In the case of the ELT, for which the redshift drift science case was first developed and is now quite robust (being a key science driver for one if its instruments), they key advantage is probing the deep matter era. While in this regime the physical mechanism responsible for the acceleration of the universe is still not dominating the dynamics, having access to a large redshift lever arm is crucial to accurately constrain this mechanism---whose dynamics is known to be slow. In any case these measurements do lead to very significant gains when they are combined with priors or with measurements from other facilities, due to the broken degeneracies. For the SKA and CHIME, for which the corresponding science cases are less developed, our results provide motivation for more detailed feasibility studies.

Given the particular redshift dependence of the drift signal (with acceleration leading to a positive drift while deceleration leads to a negative drift) there are strong advantages in combining measurements at different redshifts. It is clear that the ELT, which probes the deep matter era, is predominantly sensitive to the matter density $\Omega_m$, while the SKA is predominantly sensitive to the dark energy equation of state parameters---that is, $w_0$ and, for the CPL parametrization, $w_a$. On the other hand, CHIME or HIRAX can probe a redshift range intermediate between those of the ELT and SKA, and roughly corresponding to the end of matter domination and onset of acceleration. A practical example is the positive orientation of the ($w_0,w_a$) joint confidence contour at low redshifts, which becomes negative at high redshifts. This superficially peculiar behaviour is an illustration of the fact that the redshift drift is intrinsically different from current standard cosmological probes.

Our work should be seen as motivation for more detailed feasibility studies (supported by realistic simulations) of SKA and CHIME measurements, and we also note that synergies with other facilities should be further explored. In the present work our focus was in the redshift drift {\it per se}, so other facilities were taken into account only to the extent of providing priors on several cosmological parameters. A more thorough analysis is clearly warranted, for example in the case of Euclid \citep{Euclid}. For the case of the SKA, further possibilities include constraints on large-scale anisotropies (by measuring the drift along different directions) or measurements of the drift of the drift, enabling a cosmographic analysis \citep{Second}. We will return to these issues if future work. In any case our present conclusion is that a model-independent mapping of the expansion of the universe from redshift $z=0$ to $z=4$ is a challenging but feasible goal for the next generation of astrophysical facilities, and has an important role to play in fundamental cosmology.

\section*{Acknowledgements}

This work was financed by FEDER---Fundo Europeu de Desenvolvimento Regional funds through the COMPETE 2020---Operational Programme for Competitiveness and Internationalisation (POCI), and by Portuguese funds through FCT---Funda\c c\~ao para a Ci\^encia e a Tecnologia in the framework of the project POCI-01-0145-FEDER-028987.

ACL is supported by an FCT fellowship (SFRH/BD/113746/2015), under the FCT PD Program PhD::SPACE (PD/00040/2012). CSA and JGM were partially supported by the Scientific Initiation grants CIAAUP-06/2017-BIC and CIAAUP-06/2018-BIC, funded by FCT through national funds (UID/FIS/04434/2013) and by FEDER through COMPETE2020 (POCI-01-0145-FEDER-007672).

Several interesting discussions on the topics of this work with Joe Liske, Maria Faria and Matteo Martinelli are gratefully acknowledged.

\bibliographystyle{mnras}
\bibliography{drift} 

\appendix

\section{The impact of larger time spans}

In the main body of this article we have endeavoured to use realistic assumptions for the sensitivities of the three observational facilities under consideration, as well as for the time spans for each measurement. Here we allow ourselves to be somewhat more speculative and briefly study how constraints on model parameters would improved if one assumed obervation time spans that are larger than the baseline ones by factors of two or three. While this might larger than the expected lifetime of the instruments, it is nevertheless interesting the cosmologial impact of such observations, since one distinguishing feature of the redshift drift as a cosmological observable is that the signal grows linearly with time. 

We shold emphasize that with our assumptions (described in the main text) will benefit the ELT and CHIME, but not the SKA. The reason is that for the ELT and the SKA we are assuming absolute velocity sensitivities of the order of cm/s while for the SKA we have, following \citet{Klockner}, assumed reative errors at the percent level. The latter is clearly an approximation, but given the current uncertainties on the configuration of the SKA Phase 2 it does not seem necessary to go beyond it.

\begin{table*}
\centering
\caption{Results of the Fisher Matrix analysis for the $\Lambda$CDM model, with different choices of observation time spans. Each pair of lines shows the one-sigma marginalized uncertainties for the two parameters. The first pair corresponds to the baseline time spans discussed in the main text (and already listed in Table \ref{table1}), while the following ones correspond to larger time spans by factors of two and three respectively. The {\it All} case corresponds to the combination ELT$+$SKA$+$CHIME, while {\it C} and {\it F} respectively denote the current (Planck-like) and future (CORE-like) priors on $\Omega_mh^2$ discussed in the text.}
\label{tablea1}
\resizebox{\textwidth}{!}{%
\begin{tabular}{| c c | c c c | c c c | c c c | c c c |}
\hline
Time & Parameter & ELT & ELT+C & ELT+F & SKA & SKA+C & SKA+F & CHIME & CHIME+C & CHIME+F & All & All+C & All+F \\
\hline
Baseline & $\sigma(\Omega_m)$ &0.047 & 0.021 &  0.021 &  0.042 &  0.003 &  0.003 &  0.028 &  0.018 &  0.018 &  0.011 &  0.003  &  0.003 \\
{} & $\sigma(h)$ & 0.293  &  0.025  &  0.024  &  0.096  &  0.004  &  0.004  &  0.265  &  0.021  &  0.021  &  0.027  &  0.004  &  0.004  \\
\hline
2x Larger & $\sigma(\Omega_m)$ & 0.023  & 0.010  & 0.010  & 0.042  & 0.003  & 0.003  & 0.014  & 0.009  & 0.009  & 0.006  & 0.003  & 0.003 \\
{} & $\sigma(h)$ & 0.146  & 0.013  & 0.012  & 0.096  & 0.004  & 0.004  & 0.133  & 0.011  & 0.010  & 0.016  & 0.004  & 0.004 \\
\hline
3x Larger & $\sigma(\Omega_m)$ & 0.016  & 0.007  & 0.007  & 0.042  & 0.003  & 0.003  & 0.009  & 0.006  & 0.006  & 0.004  & 0.003  & 0.003 \\
{} & $\sigma(h)$ & 0.098  & 0.009  & 0.008  & 0.096  & 0.004  & 0.004  & 0.088  & 0.007  & 0.007  & 0.013  & 0.004  & 0.003 \\
\hline
\end{tabular}}
\end{table*}

The results are listed in Table \ref{tablea1} for the $\Lambda$CDM model, and in Table \ref{tablea2} for the CPL model. The results confirm that, for the ELT or CHIME alone, the uncertainties in the cosmological paramters are inversely proportional to the time of observations: if that time is increaed by some factor, the final uncertainty will decrease (that is, improve) by the same factor. This is still approximately true when one includes current or future priors or combines the various facilities, while if both of these are done the gains are somewhat smaller (but still clearly noticeable, espacially in the case of the CPL model.

\begin{table*}
\centering
\caption{Results of the Fisher Matrix analysis for the CPL model, with different choices of observation time spans. Each set of lines shows the one-sigma marginalized uncertainties for the four parameters. The first pair corresponds to the baseline time spans discussed in the main text (and already listed in Table \ref{table3}), while the following ones correspond to larger time spans by factors of two and three respectively. The {\it All} case corresponds to the combination ELT$+$SKA$+$CHIME, while {\it C} and {\it F} respectively denote the current (Planck-like) and future (CORE-like and Euclid-like) priors on $\Omega_mh^2$, $w_0$ and $w_a$ discussed in the text.}
\label{tablea2}
\resizebox{\textwidth}{!}{%
\begin{tabular}{| c c | c c c | c c c | c c c | c c c |}
\hline
Time & Parameter & ELT & ELT+C & ELT+F & SKA & SKA+C & SKA+F & CHIME & CHIME+C & CHIME+F & All & All+C & All+F \\
\hline
Baseline & $\sigma(\Omega_m)$ & 4.568  & 0.022  & 0.021  & 0.348  & 0.032  & 0.008  & 4.772  & 0.025  & 0.018  & 0.068  & 0.018  & 0.008\\
{} & $\sigma(h)$& 13.129 & 0.026  & 0.024  & 46.788 & 0.037  & 0.010  & 22.140 & 0.029  & 0.021  & 0.322  & 0.021  & 0.009\\
{} & $\sigma(w_0)$& 66.768 & 0.100  & 0.020  & 35.186 & 0.076  & 0.020  & 5.707  & 0.098  & 0.020  & 0.320  & 0.042  & 0.018\\
{} & $\sigma(w_a)$& 302.459 & 0.299  & 0.100  & 26.595 & 0.258  & 0.096  & 130.763 & 0.298  & 0.100  & 0.822  & 0.233  & 0.096\\
\hline
2x Larger & $\sigma(\Omega_m)$ & 2.284  & 0.013  & 0.011  & 0.348  & 0.032  & 0.008  & 2.386  & 0.018  & 0.010  & 0.041  & 0.011  & 0.006 \\
{} & $\sigma(h)$ & 6.565  & 0.015  & 0.012  & 46.788 & 0.037  & 0.010  & 11.070 & 0.022  & 0.012  & 0.184  & 0.014  & 0.008 \\
{} & $\sigma(w_0)$ & 33.384 & 0.099  & 0.020  & 35.186 & 0.076  & 0.020  & 2.853  & 0.093  & 0.020  & 0.182  & 0.026  & 0.016 \\
{} & $\sigma(w_a)$ & 151.229 & 0.296  & 0.100  & 26.595 & 0.258  & 0.096  & 65.382 & 0.294  & 0.100  & 0.600  & 0.212  & 0.094 \\
\hline
3x Larger & $\sigma(\Omega_m)$ & 1.523  & 0.010  & 0.007  & 0.348  & 0.032  & 0.008  & 1.591  & 0.016  & 0.008  & 0.033  & 0.009  & 0.005 \\
{} & $\sigma(h)$ & 4.376  & 0.012  & 0.008  & 46.788 & 0.037  & 0.010  & 7.380  & 0.019  & 0.009  & 0.140  & 0.011  & 0.006 \\
{} & $\sigma(w_0)$ & 22.256 & 0.097  & 0.020  & 35.186 & 0.076  & 0.020  & 1.902  & 0.086  & 0.020  & 0.136  & 0.020  & 0.014 \\
{} & $\sigma(w_a)$ & 100.820 & 0.292  & 0.100  & 26.595 & 0.258  & 0.096  & 43.588 & 0.289  & 0.099  & 0.519  & 0.192  & 0.092 \\
\hline
\end{tabular}}
\end{table*}

\bsp	
\label{lastpage}
\end{document}